\documentclass[a4paper,11pt]{article}

\usepackage{jheppub}
\usepackage{graphicx}
\usepackage{dcolumn}
\usepackage{bm}
\usepackage{amsmath}
\usepackage{braket}
\usepackage{slashed}    
\usepackage{epstopdf}
\usepackage{placeins}
\usepackage{multirow}
\usepackage{makecell}
\usepackage[shortlabels]{enumitem}
\usepackage{cleveref}
\usepackage{xcolor}
\usepackage[normalem]{ulem}
\usepackage[font=small, labelfont=bf, textfont={small,it}]{caption}
\usepackage{orcidlink}
\usepackage{tikz}
\usetikzlibrary{decorations.pathreplacing,arrows.meta}

\newcommand{\CP}{\mathrm{CP}}
\newcommand{\sst}{\scriptscriptstyle}
\newcommand{\nstep}{n_{\sst\rm step}}
\newcommand{\nrep}{n_{\sst\rm rep}}
\newcommand{\ii}{{\mathrm{i}}}
\newcommand{\ee}{{\mathrm{e}}}
\newcommand{\ddiff}{{\mathrm{d}}}
\newcommand{\dd}{{\sst{\mathrm{d}}}}
\newcommand{\R}{{\sst{\mathrm{R}}}}
\newcommand{\f}{{\sst{\mathrm{f}}}}
\renewcommand{\r}{{\sst{\mathrm{r}}}}
\renewcommand{\L}{{\sst{\mathrm{L}}}}
\newcommand{\PBC}{PBC}

\newcommand{\Rs}{R_{\sst\rm s}}
\newcommand{\rex}{{\sst\rm REX}}
\newcommand{\orex}{{\sst\rm OREX}}
\newcommand{\tauint}{\tau_{\sst\rm int}}

\newcommand{\beq}{\begin{eqnarray}}
\newcommand{\eeq}{\end{eqnarray}}
\newcommand{\beqnn}{\begin{eqnarray*}}
\newcommand{\eeqnn}{\end{eqnarray*}}

\title{\centering\huge Unfreezing topology using out-of-equilibrium replica exchanges}

\author[a]{Claudio Bonanno\,\orcidlink{0000-0002-1421-6510},}
\affiliation[a]{Albert Einstein Center for Fundamental Physics, Institute for Theoretical Physics, University of Bern, Sidlerstra{\ss}e 5, CH-3012 Bern, Switzerland}
\emailAdd{claudio.bonanno@unibe.ch}

\author[b]{Andrea Bulgarelli\,\orcidlink{0009-0002-2917-6125},}
\emailAdd{abulgare@uni-bonn.de}
\affiliation[b]{Transdisciplinary Research Area ``Building Blocks of Matter and Fundamental Interactions'' (TRA Matter) and Helmholtz Institute for Radiation and Nuclear Physics (HISKP), University of Bonn, Nussallee 14-16, 53115 Bonn, Germany}

\author[c]{Elia Cellini\,\orcidlink{0000-0002-5664-9752},}
\emailAdd{elia.cellini@ed.ac.uk}
\affiliation[c]{Higgs Centre for Theoretical Physics, School of Physics and Astronomy,\\
The University of Edinburgh, Edinburgh EH9 3FD, United Kingdom}

\author[d]{Roberto Dionisio\,\orcidlink{0000-0001-8021-2262},}
\affiliation[d]{Dipartimento di Fisica, Universit\`a di Pisa \& INFN,\\
Sezione di Pisa, Largo Pontecorvo 3, I-56127 Pisa, Italy}
\emailAdd{roberto.dionisio@phd.unipi.it}

\author[e]{Alessandro Nada\,\orcidlink{0000-0002-1766-5186},}
\affiliation[e]{Dipartimento di Fisica,  Universit\`a degli Studi di Torino \& INFN, Sezione di Torino, Via Pietro Giuria 1, I-10125 Turin, Italy}
\emailAdd{alessandro.nada@unito.it}

\author[e]{Dario Panfalone\,\orcidlink{0009-0007-6651-7490},}
\emailAdd{dario.panfalone@unito.it}

\author[f]{Davide Vadacchino\,\orcidlink{0000-0002-5783-5602},}
\affiliation[f]{Centre for Mathematical Sciences, University of Plymouth, Plymouth, PL4 8AA, United Kingdom}
\emailAdd{davide.vadacchino@plymouth.ac.uk}

\author[e,g]{\\Lorenzo Verzichelli\,\orcidlink{0009-0008-0825-4845}.}
\affiliation[g]{Fakult\"at f\"ur Physik, Universit\"at Bielefeld, Universit\"atsstra\ss e 25, D-33615 Bielefeld, Germany}
\emailAdd{lorenzo.verzichelli@uni-bielefeld.de}

\abstract{Parallel Tempering on Boundary Conditions (PTBC) mitigates topological freezing by exchanging configurations between replicas that interpolate between periodic and open boundaries on a small defect. We generalize PTBC by replacing the instantaneous swap with a short out-of-equilibrium protocol, in which the defect couplings of two replicas are exchanged gradually while the fields around the defect are updated. We show that the swap rate depends only on the average work spent along the out-of-equilibrium trajectory by virtue of Crooks' and Jarzynski's fluctuation relations. From the theoretical derivation of the scaling of the work as a function of the number of replicas, protocol length, and defect size, we predict the swap rate and the optimal protocol length. These predictions are tested in the $2d$ $\CP^{N-1}$ model with $N=21$. The optimal setup agrees with the measured one, the total computational cost at fixed statistical precision decreases by $\gtrsim40\%$ with respect to standard PTBC for all explored lattice spacings, and the required number of replicas at fixed acceptance is halved.
}

\begin{document}
\maketitle
\flushbottom
\section{Introduction}
\label{sec:intro}

Quantum Chromodynamics (QCD) describes the strong interaction of quarks and gluons in the Standard Model. At low energies, where the theory becomes strongly coupled, most of its physical properties can only be computed non-perturbatively. Topology is one of the clearest manifestations of such non-perturbative physics: the space of gauge field configurations splits into topological sectors, labeled by an integer topological charge $Q$, and the fluctuations of $Q$ are intimately connected with several distinctive non-perturbative features of gauge theories. Examples are the dependence of the vacuum energy on the $\theta$ parameter~\cite{Jackiw:1976pf,Vicari:2008jw}, the strong CP problem and its axion solution~\cite{Peccei:1977hh,Peccei:1977ur,Wilczek:1977pj,Weinberg:1977ma}, the anomalous breaking of the $\mathrm{U}(1)_{\mathrm{A}}$ symmetry~\cite{tHooft:1976rip} and the origin of the large mass of the $\eta^\prime$ meson~\cite{Witten:1979vv,Veneziano:1979ec,Witten:1980sp, DiVecchia:1980yfw}.

Lattice simulations are a natural first-principles tool to compute these quantities, but sampling $Q$ by Markov chain Monte Carlo (MCMC) becomes increasingly hard as the lattice spacing $a$ decreases. At finite lattice spacing, topological sectors do not exist, and they only emerge in the continuum limit~\cite{Luscher:1981zq}. At coarse lattice spacings standard local updates move freely between them; as $a\to0$ the action barriers between sectors grow and eventually diverge, and fluctuations of $Q$ become rarer and rarer. The integrated autocorrelation time of $Q^2$ grows with $1/a$ much faster than that of local observables, with a large power-law dynamical exponent or even exponentially~\cite{Alles:1996vn,DelDebbio:2002xa,DelDebbio:2004xh,Schaefer:2010hu,Bonati:2013tt,Bruno:2014ova,Bonati:2016tvi,Bonati:2017woi}, until the Markov chain remains trapped in a single sector for its whole length. This is the so-called \emph{topological freezing} problem. At fine lattice spacings it makes the calculation of the topological susceptibility and of $\theta$-dependence unfeasible, and may also affect observables that are not topological, since expectation values at fixed $Q$ differ from the correct ones by finite-volume corrections that decay with the inverse volume~\cite{Brower:2003yx,Aoki:2007ka}.

Many strategies have been proposed to mitigate topological freezing, such as open boundary conditions~\cite{Luscher:2011kk,Luscher:2012av,McGlynn:2014bxa}, metadynamics and multicanonical methods~\cite{Laio:2015era,Bonati:2018blm,Jahn:2021qrp,Eichhorn:2023uge,Eichhorn:2026eok}, density-of-states methods~\cite{Cossu:2021bgn,Borsanyi:2021gqg}, the slab method~\cite{Bietenholz:2015rsa}, master-field simulations~\cite{Luscher:2017cjh,Giusti:2018cmp,Fritzsch:2021klm}, winding transformations~\cite{Albandea:2021lvl}, and others~\cite{Funcke:2019zna,Howarth:2023bwk,Holland:2024muu,Abe:2024fpt,Holland:2025fsa,Katz:2025ycr,Blossier:2026itg}; see also Refs.~\cite{Finkenrath:2023sjg,Boyle:2024nlh} for reviews. With open boundary conditions in time, topological charge can flow in and out of the lattice, the Monte Carlo dynamics of topological modes is diffusive and the autocorrelation time of $Q^2$ grows only as $1/a^2$~\cite{Luscher:2011kk}, at the price of losing translation invariance and introducing systematic boundary effects. Parallel Tempering on Boundary Conditions (PTBC)~\cite{Hasenbusch:2017unr} combines open and periodic boundaries to keep the advantages of both choices. It simulates a set of replicas that differ only in the boundary conditions on a small defect, interpolating between periodic and open, and swaps their configurations through the replica-exchange move of parallel tempering~\cite{PhysRevLett.57.2607,Hukushima:1995mcr}. Topological excitations enter and leave through the defect of the open replicas, decorrelating the topological charge at a much faster rate. The swaps then carry the decorrelation to the periodic one, where observables are measured, free of boundary effects. Introduced for the $2d$ $\CP^{N-1}$ models~\cite{Hasenbusch:2017unr}, PTBC has since then been applied to $4d$ $\mathrm{SU}(N)$ gauge theories~\cite{Bonanno:2020hht,Bonanno:2022yjr,Bonanno:2023hhp,Bonanno:2024nba,Bonanno:2024ggk,Bonanno:2025eeb} and to QCD with dynamical fermions~\cite{Bonanno:2024zyn}. Autocorrelation times in the periodic replica are found to scale toward the continuum limit as with open boundaries~\cite{Bonanno:2025eeb}. So far, this algorithm has enabled ergodic simulations down to lattice spacings as fine as $\sim 0.02$ fm, both with and without dynamical quarks~\cite{Bonanno:2024nba,Bonanno:2025kfd}.

The computational cost of PTBC is essentially set by the number of replicas, since each replica is a full copy of the system that must be updated at every step. The replicas cannot be too few: adjacent ones exchange their configurations with an acceptance depending on the overlap between the distributions of their actions, whose fluctuations grow with the number of links crossing the defect. At fixed physical size of the defect this number grows toward the continuum limit, and the number of replicas needed to keep the acceptance fixed grows with it. Recent results in four dimensions show that the number of replicas grows with the square root of the defect volume in lattice units, i.e., $\nrep \propto \sqrt{L_\dd^3} \sim (1/a)^{3/2}$ keeping the defect size constant in physical units~\cite{Bonanno:2025eeb} (this result agrees well with general theoretical expectations about parallel tempering~\cite{Hukushima:1995mcr}). Apart from the computational overhead, the extra replicas also introduce a memory overhead. Thus, reducing the number of replicas at fixed swap acceptance rate is a natural way to make PTBC cheaper, and it is the subject of this work.

Our starting point is the observation that two distributions with different parameters can also be connected by an out-of-equilibrium transformation, rather than by an instantaneous swap between equilibrium ensembles (as done in customary parallel tempering). Jarzynski's equality~\cite{Jarzynski:1996oqb,Jarzynski:1997ef} and Crooks' theorem~\cite{Crooks:1998,Crooks_1999} relate the work done along a non-equilibrium protocol to the free-energy difference between its endpoints, and allow the unbiased computation of expectation values with respect to the final target distribution from trajectories that are never at equilibrium. After a decade of applications to free-energy calculations in lattice field theory~\cite{Caselle:2016wsw,Caselle:2018kap,Francesconi:2020fgi,Bulgarelli:2023ofi,Bulgarelli:2024onj,Bulgarelli:2025riv}, this framework was recently applied to topological freezing mitigation in Ref.~\cite{Bonanno:2024udh}: configurations drawn from an initial prior distribution with open boundaries are driven toward periodic boundary conditions by changing the defect coupling in $\nstep$ steps alternated with MCMC updates. At the end, expectation values with respect to the target distribution are recovered via an unbiased reweighting-like procedure involving the exponential of the work.

Out-of-equilibrium Monte Carlo is the purely stochastic member of the broader family of flow-based samplers, which also includes normalizing flows~\cite{Albergo:2019eim,Nicoli:2020njz,Kanwar:2020xzo,Boyda:2020hsi,DelDebbio:2021qwf,Albergo:2022qfi,Bacchio:2022,Abbott:2023thq,Albandea:2023wgd,Gerdes:2024rjk,Abbott:2024mix,Abbott:2026ylv,Abbott:2026vui,Singha:2026aac}, see Refs.~\cite{Albergo:2021vyo,Cranmer:2023xbe,Cheng:2025khv} for reviews, and diffusion models~\cite{Wang:2023exq,Zhu:2025pmw,Vega:2025hgz,Aarts:2026zzr,Alharazin:2026lcb,Komijani:2026lan,Tomiya:2026tbc,Kanwar:2026prt}. The scaling of training costs has so far prevented normalizing flows from being applied to state-of-the-art, large-scale four-dimensional calculations~\cite{DelDebbio:2021qwf,Abbott:2022zsh}. Stochastic normalizing flows~\cite{Caselle:2022acb,Bulgarelli:2024brv}, which alternate out-of-equilibrium MCMC updates with trained deterministic layers, preserve the scaling toward the continuum of out-of-equilibrium Monte Carlo at a fraction of its sampling cost and with negligible training cost. They have recently been shown to reduce the dynamical power-law exponent of the autocorrelation time of $Q^2$ in $\mathrm{SU}(3)$ Yang--Mills theory down to $a\simeq0.03$~fm~\cite{Bonanno:2025pdp,Bonanno:2026zqy}; see also Refs.~\cite{Albergo:2024trn,Chen:2026bqw} for related architectures.

Learned transformations have also been combined with parallel tempering with the purpose of reducing the number of replicas at fixed swap acceptance rate: in particular, they have been employed as maps between the distributions of adjacent replicas~\cite{Invernizzi:2022aaa,Zhang:2025aaa}, which has also been explored in lattice field theory~\cite{Abbott:2024mix,Kanwar:2026prt}. Given that out-of-equilibrium transformations can be seen as a limiting case of stochastic normalizing flows, which are also learned transformations, it is natural to explore how their combination with parallel tempering would perform in terms of replica and cost reduction.

In this work we develop this idea in its simplest form and apply it to PTBC: we replace the instantaneous swap of two configurations drawn from distributions with different boundary conditions on the defect with an out-of-equilibrium transformation, in the same spirit as the replica exchange with non-equilibrium switches of Ref.~\cite{Ballard:2009,Ballard:2012rex}, see also Ref.~\cite{Nilmeier_2011}. The defect couplings of two adjacent replicas are exchanged gradually, in $\nstep$ steps, and each step is followed by MCMC updates of the fields around the defect. The move is accepted with a Metropolis probability depending on the work done along the out-of-equilibrium protocol. By Crooks' theorem the move satisfies detailed balance for any $\nstep$, so the algorithm is exact, and for $\nstep=0$ it reduces to the standard swap move. We call this algorithm Out-of-equilibrium Replica Exchange (O-REX), as opposed to ordinary Replica Exchange (REX) parallel tempering.

Two features of PTBC make this extension promising and worthwhile. First, the average work decreases with the length of the protocol and vanishes in the quasi-static limit, so at fixed acceptance a longer protocol allows fewer replicas. Second, since only the defect coupling changes along the protocol, the system is driven out of equilibrium only near the defect, and the updates along the protocol can be restricted to a small patch around it, a few percent of the lattice volume in our simulations. A protocol then costs only a small fraction of a replica; thus, reducing the number of replicas decreases the computational cost more than the protocol increases it with respect to the instantaneous swap. The updates around the defect also decorrelate the topological charge on their own, which adds to the gain.

We test our approach in $2d$ $\CP^{N-1}$ models, focusing on $N=21$ and five lattice spacings to study the scaling of the algorithm towards the continuum limit. We chose this class of quantum field theories as our test-bed because they share many crucial features with QCD such as asymptotic freedom, the existence of instanton configurations, a non-trivial $\theta$-dependence, and a $1/N$ expansion of expectation values in the large-$N$ limit~\cite{DAdda:1978vbw,Witten:1978bc,David:1983zx}. Due to the existence of non-trivial topological properties, standard simulations of $2d$ $\CP^{N-1}$ models are also affected by a severe critical slowing down of topological modes toward the continuum limit, whose severity is as bad as the one observed in $4d$ gauge theories~\cite{Campostrini:1992ar,DelDebbio:2004xh}. For this reason, $2d$ $\CP^{N-1}$ models have often been employed as a testing ground for lattice methods~\cite{Campostrini:1992it,Vicari:1992jy,Rossi:1993nz,Laio:2015era,Flynn:2015uma,Hasenbusch:2017unr,Bonanno:2024udh}.

The paper is organized as follows. Sec.~\ref{sec:theoretical_setup} defines the theory in the continuum and on the lattice (including the defect), as well as the observables of interest. Sec.~\ref{sec:orex} introduces the new algorithm: we briefly summarize the main theoretical properties of replica exchange and out-of-equilibrium transformations, we define the O-REX general algorithmic framework, and its application to PTBC. Sec.~\ref{sec:results} presents the numerical results obtained with PTBC and Out-of-equilibrium PTBC (O-PTBC). Finally, in Sec.~\ref{sec:conclusions} we draw our conclusions and discuss the prospects for applying this framework to $4d$ gauge theories. Additional data and information are reported in the Appendices.

\section{Continuum and lattice formulations of \texorpdfstring{$2d$}{2d} \texorpdfstring{$\CP^{N-1}$}{CP(N-1)} models}\label{sec:theoretical_setup}

In this section we introduce the continuum and lattice formulations of $2d$ $\CP^{N-1}$ models, including the definition of the defect.

\subsection{The theory in the continuum}

We consider the following continuum formulation of $2d$ $\CP^{N-1}$ models:
\begin{equation}
    S[z,\bar{z},A]=\frac{N}{g}\int \ddiff^2 x\,\bar{D}_\mu \bar{z}(x)D_\mu z(x),
    \label{eq:cpn_continuum_action}
\end{equation}
where $g\sim\mathcal{O}(N^0)$ is the 't Hooft coupling, $z(x)$ is an $N$-component complex scalar field satisfying $\bar z(x)z(x)=1$, and $A_\mu(x)$ is a non-propagating auxiliary $\mathrm{U}(1)$ gauge field. The covariant derivative is $D_\mu=\partial_\mu+\ii A_\mu$. In this formulation, the action is invariant under local $\mathrm{U}(1)$ transformations:
\begin{equation}
    z(x)\longrightarrow \ee^{\ii\alpha(x)}z(x), \qquad A_\mu(x)\longrightarrow A_\mu(x)-\partial_\mu\alpha(x).
\end{equation}

Our main quantity of interest will be the topological susceptibility. The topological charge can be simply expressed in terms of the auxiliary gauge field as:
\begin{equation}
    Q=\frac{1}{2\pi}\varepsilon_{\mu\nu}\int \ddiff^2 x\,\partial_\mu A_\nu(x)\in\mathbb{Z},
    \label{eq:cpn_continuum_topological_charge}
\end{equation}
with $\varepsilon_{01} = -\varepsilon_{10}=1$ and $\varepsilon_{00}=\varepsilon_{11}=0$. The topological susceptibility reads:
\beq
\chi = \lim_{V\to\infty}\frac{\braket{Q^2}}{V}, \qquad \qquad V=\ell^2,
\eeq
where $\ell$ is the linear size of the system. In $2d$ $\CP^{N-1}$ models at large $N$, it is customary to express all dimensionful quantities in terms of the second-moment correlation length $\xi$, defined as:
\beq
\xi^2 = \frac{1}{4}\frac{\int \ddiff^2 x \, \vert x \vert^2 \, G(x)}{\int \ddiff^2 x \, G(x)},
\eeq
with $G(x)$ the connected two-point function
\begin{equation}
G(x)=\left\langle P_{ab}(x)P_{ab}(0) \right\rangle-\frac{1}{N} = \left\langle |z(x)z(0)|^2 \right\rangle-\frac{1}{N}, \qquad \qquad P_{ab}(x) \equiv z_a(x) \, \bar{z}_b(x).
    \label{eq:proj_correlator}
\end{equation}

\subsection{The theory on the lattice}

We discretize the theory on a square lattice of size $L\times L$ and lattice spacing $a$ using the tree-level Symanzik-improved action. As usual, the scalar field $z(x)$ lives on the site $x$, while the $\mathrm{U}(1)$ gauge variable $U_\mu(x)\simeq\exp\{\ii a A_\mu(x)\}$ lives on the link between sites $x$ and $x+a\hat{\mu}$. The lattice action reads~\cite{Campostrini:1992ar}:
\begin{equation}
\begin{aligned}
    S_{\L}^{\sst{(c)}}[z,\bar z,U]&=-2N\beta\sum_{x,\mu}\Big\{C^{\sst\rm (Sym)}_1 k^{\sst{(c)}}_\mu(x)\,\mathrm{Re}\left[\bar U_\mu(x)\,\bar z(x+a\hat\mu)z(x)\right]  +\\
    &+C^{\sst\rm (Sym)}_2 k^{\sst{(c)}}_\mu(x)k^{\sst{(c)}}_\mu(x+a\hat\mu) \,\mathrm{Re}\left[\bar U_\mu(x+a\hat\mu)\bar U_\mu(x)\,\bar z(x+2a\hat\mu)z(x)\right]\Big\},
    \label{eq:cpn_lattice_action}
\end{aligned}
\end{equation}
where $\beta=1/g_0$ is the inverse bare 't Hooft coupling and $C^{\sst\rm (Sym)}_1=4/3$, $C^{\sst\rm (Sym)}_2=-1/12$ are the Symanzik-improvement coefficients.

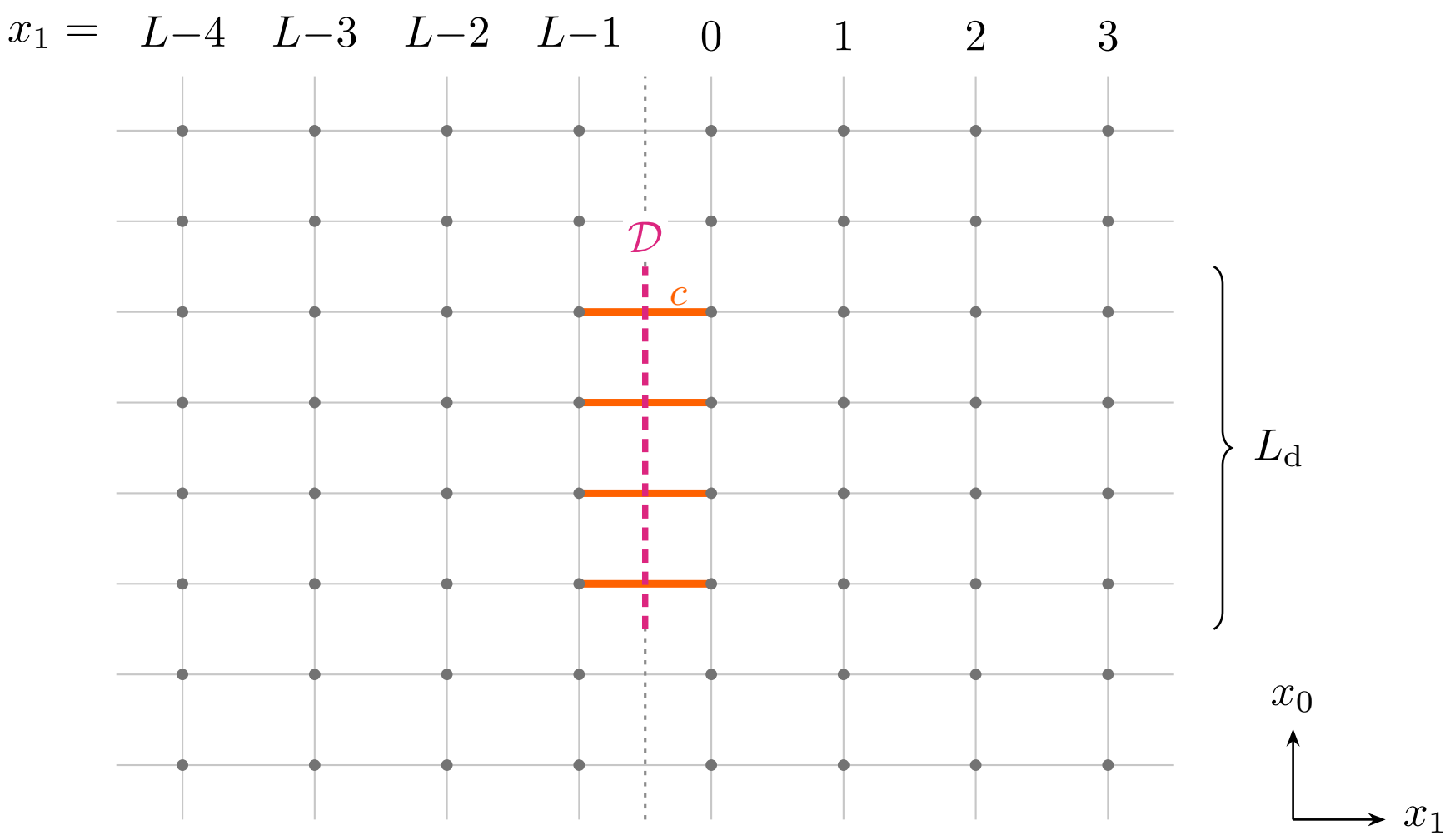
\begin{figure}[t]
  \centering
\providecolor{ibmmagenta}{HTML}{DC267F}
\providecolor{ibmorange}{HTML}{FE6100}
\begin{tikzpicture}[x=1.05cm, y=0.72cm,
    site/.style={circle, fill=black!55, inner sep=0pt, minimum size=2.6pt},
    gridline/.style={black!22, line width=0.4pt},
    boundary/.style={black!45, line width=0.6pt, dash pattern=on 1pt off 1.5pt},
    defect/.style={ibmmagenta, line width=1.4pt, dash pattern=on 3pt off 2pt},
    nn/.style={ibmorange, line width=1.7pt},
    brace/.style={decorate, decoration={brace, amplitude=4pt}, line width=0.5pt},
    lab/.style={font=\small},
  ]
  \foreach \i in {0,...,7} \draw[gridline] (\i,-0.6) -- (\i,7.6);
  \foreach \j in {0,...,7} \draw[gridline] (-0.5,\j) -- (7.5,\j);
  \draw[boundary] (3.5,-0.6) -- (3.5,1.5);
  \draw[boundary] (3.5,5.5) -- (3.5,7.6);
  \foreach \j in {2,...,5} \draw[nn] (3,\j) -- (4,\j);
  \draw[defect] (3.5,1.5) -- (3.5,5.5);
  \foreach \i in {0,...,7} \foreach \j in {0,...,7} \node[site] at (\i,\j) {};
  \foreach \i/\t in {0/{L{-}4},1/{L{-}3},2/{L{-}2},3/{L{-}1},4/{0},5/{1},6/{2},7/{3}}
    \node[lab, anchor=south] at (\i,7.7) {$\t$};
  \node[lab, anchor=east] at (-0.5,8.05) {$x_1 =$};
  \node[lab, text=ibmmagenta, anchor=south, fill=white, inner sep=1pt] at (3.5,5.6) {$\mathcal{D}$};
  \node[lab, text=ibmorange, anchor=south, inner sep=1.5pt] at (3.75,5) {\footnotesize $c$};
  \draw[brace, decoration={mirror}] (7.8,1.5) -- node[lab, right=5pt] {$L_{\rm d}$} (7.8,5.5);
  \draw[-{Stealth[length=4pt]}, line width=0.5pt] (8.4,-0.6) -- (9.1,-0.6) node[lab, right] {$x_1$};
  \draw[-{Stealth[length=4pt]}, line width=0.5pt] (8.4,-0.6) -- (8.4,0.4) node[lab, above] {$x_0$};
\end{tikzpicture}
  \caption{Lattice around the defect $\mathcal{D}$ (dashed), located on the spatial
  boundary between $x_1 = L-1$ and $x_1 = 0$ and extending over $L_{\rm d}$ sites in
  the temporal direction, cf.~Eq.~\eqref{eq:cpn_defect_def}. The couplings of the links in the $\mu=1$ direction crossing the defect (orange) are weighted by $c$, cf.~Eq.~\eqref{eq:cpn_defect}.}
  \label{fig:defect}
\end{figure}

The factors $k^{\sst{(c)}}_\mu(x)$ modify the couplings across a segment of the spatial boundary, defining a defect of temporal extent $L_\dd$ lattice spacings (see Fig.~\ref{fig:defect}). Writing the lattice sites as $x=(x_0,x_1)$, where $x_0$ and $x_1$ are the temporal and spatial coordinates and $x_\mu=0,\ldots,L-1$, we identify this segment with
\begin{equation}
    \mathcal{D}=\left\{x \,:\, x_1 = L-1,\ 0\le x_0 < L_\dd \right\}, \qquad 1\le L_\dd\le L,
    \label{eq:cpn_defect_def}
\end{equation}
and set
\begin{equation}
    k^{\sst{(c)}}_\mu(x)=
    \begin{cases}
        c, & x\in\mathcal{D}\ \text{and}\ \mu=1,\\
        1, & \text{otherwise},
    \end{cases}
    \label{eq:cpn_defect}
\end{equation}
with $c\in[0,1]$. Varying $c$ interpolates between periodic boundary conditions (\PBC) at $c=1$ and open boundary conditions (OBC) along the defect at $c=0$. For $L_\dd=L$, the defect covers the whole spatial boundary; for $L_\dd<L$, the boundary remains periodic outside the defect.

For future convenience, it is useful to recall that, since $x$ and $x+a\hat 1$ cannot both belong to $\mathcal{D}$, the product $k^{(c)}_1(x)\,k^{(c)}_1(x+a\hat 1)$ in the next-to-nearest-neighbor term contains at most one power of $c$. The action is therefore linear in $c$:
\begin{equation}
    S_{\L}^{\sst{(c)}}[z,\bar z,U]=S_{\L}^{\sst{(\rm\PBC)}}[z,\bar z,U]+c\,S_{\L}^{\sst{(\mathcal{D})}}[z,\bar z,U],
    \label{eq:cpn_action_linear_c}
\end{equation}
where $S_{\L}^{\sst{(\mathcal{D})}}$ collects the terms in the sum defining the right-hand side of Eq.~\eqref{eq:cpn_lattice_action} involving links that cross the defect, and $S_{\L}^{\sst{(\rm\PBC)}}$ the remaining ones. When changing the parameter $c$ to modify the boundary conditions along the defect, only $S_{\L}^{\sst{(\mathcal{D})}}$ will matter to compute the total action difference.

Let us now discuss the main observables we will be interested in. We recall that they will always be evaluated on the periodic replica with $c=1$, where translation invariance is restored. Concerning the topological charge, several lattice definitions are possible in $2d$ $\CP^{N-1}$ models~\cite{Berg:1981er,Campostrini:1992ar}. In this work, we discretize it via the following geometric formulation, defined in terms of the gauge fields~\cite{Campostrini:1992ar}:
\begin{equation}
    Q_{\sst\rm geo}[U]=\frac{1}{2\pi}\sum_x\mathrm{Im}\left\{\log\left[U_{01}(x)\right]\right\}\in\mathbb{Z},
    \label{eq:cpn_geometric_topological_charge}
\end{equation}
where
\begin{equation}
    U_{\mu\nu}(x) \equiv U_\mu(x)U_\nu(x+a\hat\mu)\bar U_\mu(x+a\hat\nu)\bar U_\nu(x)
    \label{eq:cpn_plaquette}
\end{equation}
is the plaquette, and the natural logarithm is evaluated on its principal branch. With \PBC, this definition gives an integer-valued charge for every lattice configuration. Concerning the topological susceptibility, it is customary to evaluate it in terms of smoothened fields to suppress UV fluctuations that could introduce multiplicative and additive renormalizations in the lattice topological susceptibility~\cite{Alles:1997nu}. Several smoothing algorithms have been adopted to this purpose, such as cooling~\cite{Berg:1981nw,Iwasaki:1983bv,Itoh:1984pr,Teper:1985rb,Ilgenfritz:1985dz,Campostrini:1989dh,Alles:2000sc}, stout smearing~\cite{APE:1987ehd, Morningstar:2003gk} or gradient flow~\cite{Narayanan:2006rf,Luscher:2010iy}. All these methods give agreeing results once the smoothing radii are matched among them~\cite{Alles:2000sc, Bonati:2014tqa, Alexandrou:2015yba}. In this work we choose cooling for its computational cheapness, following the implementation given in~\cite{Alles:2000sc}. In particular, the action that is minimized during cooling sweeps is the unimproved one with $C^{\sst\rm (Sym)}_1=1$ and $C^{\sst\rm (Sym)}_2=0$. The topological susceptibility is defined as:
\begin{equation}
\chi = \frac{1}{V}\braket{Q^2_{\sst\rm geo}[U_{\sst\rm cool}(\Rs)]}, \qquad \qquad V=\ell^2=(aL)^2,
    \label{eq:lat_top_susc}
\end{equation}
where $U_{\sst\rm cool}(\Rs)$ stands for the gauge fields obtained after $n_{\sst\rm cool}$ cooling steps, and $\Rs$ is the smoothing radius, with $\Rs^2/a^2 \propto n_{\sst\rm cool}$, below which UV fluctuations are damped. In the continuum limit, $\chi$ will be independent of $\Rs$~\cite{Ce:2015qha,Bonanno:2025eeb,Durr:2025qtq}.

Concerning the second-moment correlation length, we adopt the following lattice discretization~\cite{Caracciolo:1998gga}:
\begin{equation}
\left(\frac{\xi}{a}\right)^2=\frac{1}{4\sin^2(ap_{\sst\rm min}/2)}\left[\frac{\chi_{_{\sst\rm m}}}{\widetilde G_{\L}(p_{\sst\rm min})}-1\right], \qquad ap_{\sst\rm min}=\frac{2\pi}{L}.
    \label{eq:xi_sec_mom}
\end{equation}
Here $\widetilde{G}_{\L}(p)=\sum_x \exp(\ii p_\mu x_\mu)\, G_{\L}(x)$, with $G_{\L}(x)$ the lattice counterpart of Eq.~\eqref{eq:proj_correlator}. The quantity $\chi_{_{\sst\rm m}}$ is the so-called magnetic susceptibility and coincides with the zero-momentum correlator:
\begin{equation}
    \chi_{_{\sst\rm m}}=\widetilde{G}_{\L}(0)=\sum_x G_{\L}(x).
    \label{eq:magn_susc}
\end{equation}
Finally, we also computed the action density, defined as:
\begin{equation}
    \begin{split}
        a^2E=\frac{1}{L^2}\sum_{x,\mu}\Big\{&C^{\sst{\rm (Sym)}}_1\left(1-\mathrm{Re}\left[\bar U_\mu(x)\,\bar z(x+a\hat\mu)z(x)\right]\right)\\
        &+C^{\sst{\rm (Sym)}}_2\left(1-\mathrm{Re}\left[\bar U_\mu(x+a\hat\mu)\bar U_\mu(x)\,\bar z(x+2a\hat\mu)z(x)\right]\right)\Big\},
    \end{split}
    \label{eq:action_density}
\end{equation}
or equivalently $a^2E=2(C^{\sst{\rm (Sym)}}_1+C^{\sst{\rm (Sym)}}_2)+S^{\sst{(c\,=\, 1)}}_{\L}[z,\bar z,U]/(2N\beta L^2)$. 

\subsection{Standard algorithms and autocorrelation times}

One of the benefits of adopting the gauge field formulation of $2d$ $\CP^{N-1}$ models is that the lattice action is local in both fields $U_\mu(x)$ and $z(x)$. This allows one to use standard local updating algorithms such as heat-bath and over-relaxation~\cite{Campostrini:1992ar}. More details about the parallel tempering updating scheme will be given later in Sec.~\ref{sec:results}.

Error estimation of expectation values is performed via a standard binned jack-knife analysis. To monitor the decorrelation of the topological charge, we also computed its integrated autocorrelation time via the customary Gamma method~\cite{Wolff:2003sm}. For a generic observable $\mathcal{O}[z,\bar{z},U]$ measured along the Markov chain, the integrated autocorrelation time is defined as:
\begin{equation}
    \tauint(\mathcal{O})=\frac{1}{2}+\sum_{t\,=\,1}^{\infty}\frac{\Gamma_{\mathcal{O}}(t)}{\Gamma_{\mathcal{O}}(0)}, \qquad \Gamma_{\mathcal{O}}(t)=\left\langle\mathcal{O}_s\,\mathcal{O}_{s+t}\right\rangle-\left\langle\mathcal{O}\right\rangle^2,
    \label{eq:tau_int}
\end{equation}
where $\mathcal{O}_s$ is the measurement at Monte Carlo time $s$. In practice, the sum is truncated at a finite window $W$, chosen sufficiently large that the missing contribution from the tail of $\Gamma_{\mathcal{O}}(t)$ at large $t$ can be neglected. The Monte Carlo time unit and how to take into account the extra computational cost of the replicas in the two algorithms (REX and O-REX) will be discussed later in Sec.~\ref{sec:results}.


\section{Out-of-equilibrium replica exchange}
\label{sec:orex}

In this section we introduce the Out-of-equilibrium Replica Exchange (O-REX)~\cite{Ballard:2009,Ballard:2012rex} algorithm discussed in the introduction. We first review the main features of the Replica Exchange (REX) algorithm~\cite{PhysRevLett.57.2607}, and of the out-of-equilibrium framework based on Jarzynski's~\cite{Jarzynski:1996oqb,Jarzynski:1997ef} and Crooks'~\cite{Crooks:1998,Crooks_1999} theorems. We then combine the two and define the O-REX, in which the instantaneous swap of configurations is replaced by a gradual out-of-equilibrium transformation. Finally, we discuss the practical application of the O-REX framework to Parallel Tembering on Boundary Conditions (PTBC).

\subsection{Standard replica exchange}
\label{subsec:rex}

Consider a family of target Boltzmann distributions labeled by a set of parameters $\theta$,
\begin{equation}
    p_\theta(x)=\frac{1}{Z_\theta}\ee^{-S_\theta(x)},\qquad Z_\theta=\int \ddiff x\, \ee^{-S_\theta(x)},
    \label{eq:boltzmann_theta}
\end{equation}
where $x$ denotes a field configuration and $S_\theta$ the dimensionless action. The parameter $\theta$ may be, for instance, the inverse coupling $\beta$, as in standard parallel tempering, or the defect coupling $c$ of Sec.~\ref{sec:theoretical_setup}. 
In parallel tempering, one simulates an extended ensemble composed of several replicas at different values of $\theta$. For two replicas with configurations $x_i,x_j$ and parameters $\theta_i,\theta_j$, and using the shorthand $S_i\equiv S_{\theta_i}$, the joint target distribution factorizes as:
\begin{equation}
    \Pi_{ij}(x_i,x_j)=p_{\theta_i}(x_i)\,p_{\theta_j}(x_j)=\frac{1}{Z_{ij}}\exp\left\{-S_{ij}(x_i,x_j)\right\},
    \label{eq:two_replica_ensemble}
\end{equation}
\begin{equation}
    \hspace{-1.8\baselineskip} S_{ij}(x_i,x_j)=S_i(x_i)+S_j(x_j), \qquad Z_{ij}=Z_{\theta_i}Z_{\theta_j},
\end{equation}
where $S_{ij}$ is the action of the two-replica extended ensemble.

A REX move proposes to swap the two configurations while keeping the parameters fixed, i.e., $X\to X'$ with $X=(x_i,x_j)$ and $X'=(x_j,x_i)$. Assuming a symmetric proposal probability, detailed balance with respect to $\Pi_{ij}$,
\begin{equation}
\label{eq:rex_detbal}
    \Pi_{ij}(X)\,A_{\rex}(X\to X')=\Pi_{ij}(X')\,A_{\rex}(X'\to X),
\end{equation}
is satisfied by the standard Metropolis acceptance probability
\begin{equation}
    A_{\rex}(X\to X')=\min\left\{1,\ee^{-\Delta_{ij} (X,X') }\right\},
    \label{eq:rex_acceptance}
\end{equation}
where
\begin{equation}
    \Delta_{ij} (X,X') \equiv S_{ij}(X')-S_{ij}(X)=\left[S_j(x_i)-S_i(x_i)\right]-\left[S_j(x_j)-S_i(x_j)\right].
    \label{eq:rex_delta}
\end{equation}
For $S_\theta(x)=\beta E(x)$, this reduces to the familiar expression $\Delta_{ij}=-(\beta_i-\beta_j)\left[E(x_i)-E(x_j)\right]$.

\subsection{Out-of-equilibrium transformations}
\label{subsec:noneq}

Consider now a sequence of actions $S_0,S_1,\ldots,S_{\nstep}$, with associated normalized Boltzmann densities $p_n(x)=\ee^{-S_n(x)}/Z_n$, and free-energy difference $\Delta F=-\log(Z_{\nstep}/Z_0)$ between the final and initial distributions.
A non-equilibrium trajectory $\mathcal{X}=(x_0,x_1,\ldots,x_{\nstep})$ is generated by a forward protocol made of a sequence of MCMC updates, starting from $x_0\sim p_0$: at each step the action is changed from $S_{n-1}$ to $S_n$, and the configuration is then updated with a transition kernel $P_n$ that leaves $p_n$ unchanged. The full forward path probability density reads
\begin{equation}
    p_0(x_0)\,\mathcal{P}_\f(\mathcal{X}),\qquad \mathcal{P}_\f(\mathcal{X})=\prod_{n=1}^{\nstep}P_n(x_{n-1}\to x_n).
\end{equation}
Analogously, the time-reversed trajectory $\tilde{\mathcal{X}}=(x_{\nstep},x_{\nstep-1},\ldots,x_0)$ has full path probability density
\begin{equation}
    p_{\nstep}(x_{\nstep})\,\mathcal{P}_\r(\tilde{\mathcal{X}}),\qquad \mathcal{P}_\r(\tilde{\mathcal{X}})=\prod_{n=1}^{\nstep}\widetilde{P}_n(x_n\to x_{n-1}),
\end{equation}
where $\widetilde{P}_n$ is the transition kernel associated with the reverse protocol.
Defining the work performed along the trajectory as
\begin{equation}
    W(\mathcal{X})=\sum_{n=0}^{\nstep-1}\left[S_{n+1}(x_n)-S_n(x_n)\right],
    \label{eq:work}
\end{equation}
Crooks' fluctuation theorem~\cite{Crooks:1998,Crooks_1999} relates the forward and reverse path probabilities,
\begin{equation}
    \frac{p_0(x_0)\,\mathcal{P}_\f(\mathcal{X})}{p_{\nstep}(x_{\nstep})\,\mathcal{P}_\r(\tilde{\mathcal{X}})}=\exp\left[W(\mathcal{X})-\Delta F\right].
    \label{eq:crooks}
\end{equation}
In general, Eq.~\eqref{eq:crooks} holds provided the dynamics is microscopically reversible, i.e.,
\beq
p_n(x)\,P_n(x\to y)=p_n(y)\,\widetilde{P}_n(y\to x)
\eeq
for every $n$ and every $x$,$y$. If $P_n$ satisfies detailed balance with respect to $p_n$, this condition is fulfilled with $\widetilde{P}_n=P_n$, i.e., if $P_n$ is a certain sequence of updates, $\widetilde{P}_n$ is the same sequence applied in reverse order. Note that in an out-of-equilibrium transformation the reverse path is not computed explicitly, and $\mathcal{P}_\r(\tilde{\mathcal{X}})$ does not have to be implemented in practice: $\widetilde{P}_n=P_n$ is just a formal statement about the possibility to go through the forward path exactly in reverse order. As we shall see later, this is not the case for O-REX, and the reverse path will be an important ingredient of the algorithm.

As a direct consequence of Eq.~\eqref{eq:crooks}, one obtains Jarzynski's equality~\cite{Jarzynski:1996oqb,Jarzynski:1997ef}
\begin{equation}
    \left\langle \ee^{-W}\right\rangle_\f=\ee^{-\Delta F},
    \label{eq:jarzynski}
\end{equation}
where the expectation value is taken over forward non-equilibrium trajectories.
The same framework underlies several algorithmic proposals: annealed importance sampling~\cite{Neal2001} and its applications to lattice field theory~\cite{Caselle:2016wsw}, the out-of-equilibrium Monte Carlo approach to topological freezing of Ref.~\cite{Bonanno:2024udh}, and stochastic normalizing flows~\cite{Caselle:2022acb,Bulgarelli:2024brv,Bonanno:2025pdp,Bonanno:2026zqy}.

\subsection{Out-of-equilibrium replica exchange}
\label{subsec:orex}

\begin{figure}[!t]
\centering
\includegraphics[width=0.9\textwidth]{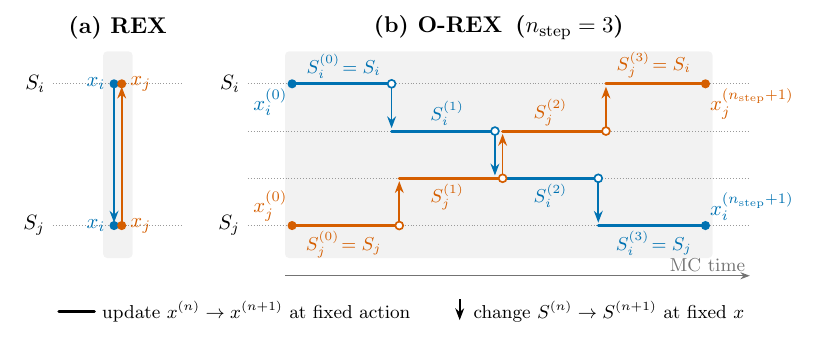}
\caption{Exchange move between two adjacent replicas. (a) In REX, the swap is instantaneous. (b) In O-REX the exchange is gradual: $\nstep$ changes of
  the actions at fixed configuration (vertical arrows) alternate with $\nstep+1$
  updates of the configurations at fixed action (horizontal segments), so that the
  replica starting from $S_i$ ends with $S_i^{(\nstep)}=S_j$, and vice versa.
}
\label{fig:orex_schematic}
\end{figure}

We now generalize standard replica exchange by replacing the instantaneous swap with a finite out-of-equilibrium transformation, along the lines of Ref.~\cite{Ballard:2009,Ballard:2012rex}. Consider two replicas initially assigned to parameters $\theta_i$ and $\theta_j$, with initial extended equilibrium state $X_0=(x_{i}^{\sst{(0)}},x_{j}^{\sst{(0)}})$ distributed according to $\Pi_{ij}^{\sst{(0)}}\equiv\Pi_{ij}$ of Eq.~\eqref{eq:two_replica_ensemble}. In O-REX, the two replicas are driven out of equilibrium from $S_i$ to $S_j$ and from $S_j$ to $S_i$, respectively. This is achieved by alternating $\nstep$ steps in which $\lambda_n$ is changed at fixed field configuration, with $\nstep+1$ updates of the configuration at fixed value of $\lambda_n$:
\beq
\begin{aligned}
\underbrace{S_i^{\sst{(0)}}(x_i^{\sst{(0)}})}_{\equiv \,S_i(x_i)} \to S_i^{\sst{(0)}}(x_i^{\sst{(1)}}) \to S_i^{\sst{(1)}}(x_i^{\sst{(1)}}) & \to \dots \\
&\to S_i^{\sst{(\nstep)}}(x_i^{\sst{(\nstep)}}) \to \underbrace{S_i^{\sst{(\nstep)}}(x_i^{\sst{(\nstep+1)}})}_{\equiv \,S_j(x_i^{\sst{(\nstep+1)}})}
\end{aligned}
\eeq
\beq
\begin{aligned}
\underbrace{S_j^{\sst{(0)}}(x_j^{\sst{(0)}})}_{\equiv \,S_j(x_j)} \to S_j^{\sst{(0)}}(x_j^{\sst{(1)}}) \to S_j^{\sst{(1)}}(x_j^{\sst{(1)}}) & \to \dots \\
&\to S_j^{\sst{(\nstep)}}(x_j^{\sst{(\nstep)}}) \to \underbrace{S_j^{\sst{(\nstep)}}(x_j^{\sst{(\nstep+1)}})}_{\equiv \,S_i(x_j^{\sst{(\nstep+1)}})}
\end{aligned}
\eeq
where an increase in the superscript of $x_i^{\sst{(n)}}$ denotes an updating step of the configuration at fixed defect coupling, while a change in the superscript of $S^{\sst{(n)}}$ denotes a change in $\lambda_n$ at fixed field configuration. The protocol is sketched in Fig.~\ref{fig:orex_schematic}, where it is also compared to the instantaneous swap of standard REX. 

For a linear protocol, one can write:
\beq
S_i^{\sst{(m)}}(x_i^{\sst{(n)}}) &=& (1-\lambda_m)S_i(x_i^{\sst{(n)}})+\lambda_m S_j(x_i^{\sst{(n)}}),\\
S_j^{\sst{(m)}}(x_j^{\sst{(n)}}) &=& (1-\lambda_m)S_j(x_j^{\sst{(n)}})+\lambda_m S_i(x_j^{\sst{(n)}}),
\eeq
with $\lambda_m=m/\nstep$ and $m=0,\ldots,\nstep$. The joint action thus reads:
\beq
\begin{aligned}
S_{ij}^{\sst{(m)}}(x_i^{\sst{(n)}},x_j^{\sst{(n)}}) &= S_i^{\sst{(m)}}(x_i^{\sst{(n)}})+S_j^{\sst{(m)}}(x_j^{\sst{(n)}}) \\
&= (1-\lambda_m)\left[S_i(x_i^{\sst{(n)}})+S_j(x_j^{\sst{(n)}})\right]+\lambda_m\left[S_j(x_i^{\sst{(n)}})+S_i(x_j^{\sst{(n)}})\right].
\label{eq:orex_action}
\end{aligned}
\eeq
When $S_\theta$ depends linearly on $\theta$, a linear protocol amounts to interpolating linearly between the two parameters. Note that, since the protocol exchanges $S_i$ and $S_j$ between the two replicas, the initial and final partition functions coincide, $Z_0=Z_{\theta_i}Z_{\theta_j}=Z_{\nstep}$. Therefore:
\begin{equation}
    \Delta F=0 .
\end{equation}
The exchange protocol generates a non-equilibrium trajectory $\Gamma=(X_0,\ldots,X_{\nstep+1})$, with $X_n=(x_{i}^{\sst{(n)}},x_{j}^{\sst{(n)}})$, by alternating changes of the joint action with MCMC updates of both replicas. Its forward and reverse path probability densities:
\beq
\Pi_{ij}^{\sst{(0)}}(X_0)\,\mathcal{P}_\f(\Gamma), \qquad &\text{(forward probability)},\\
\Pi_{ij}^{\sst{(\nstep)}}(X_{\nstep+1})\,\mathcal{P}_\r(\tilde{\Gamma}), \qquad &\text{(reverse probability)},
\eeq
are defined as in Sec.~\ref{subsec:noneq}. Here $\tilde{\Gamma}=(X_{\nstep+1}, \dots, X_0)$ is the reverse trajectory. Using Eq.~\eqref{eq:orex_action}, the work of Eq.~\eqref{eq:work} takes the simple form:
\begin{equation}
    \begin{split}
    W_{ij}(\Gamma)&=\sum_{n=1}^{\nstep}\left[S_{ij}^{\sst{(n)}}(X_n)-S_{ij}^{\sst{(n-1)}}(X_n)\right]\\
    &=\frac{1}{\nstep}\sum_{n=1}^{\nstep}\Big\{\left[S_j(x_{i}^{\sst{(n)}})-S_i(x_{i}^{\sst{(n)}})\right]-\left[S_j(x_{j}^{\sst{(n)}})-S_i(x_{j}^{\sst{(n)}})\right]\Big\}.
    \end{split}
    \label{eq:orex_work}
\end{equation}
Using Crooks' theorem~\eqref{eq:crooks} with $\Delta F=0$ finally yields:
\begin{equation}
    \frac{\Pi_{ij}^{\sst{(0)}}(X_0)\,\mathcal{P}_\f(\Gamma)}{\Pi_{ij}^{\sst{(\nstep)}}(X_{\nstep+1})\,\mathcal{P}_\r(\tilde{\Gamma})}=\ee^{W_{ij}(\Gamma)}.
    \label{eq:orex_crooks}
\end{equation}
This formula has the same role as the relation $\Pi_{ij}(X)/\Pi_{ij}(X')=\ee^{\Delta_{ij}(X,X')}$ in the standard replica exchange, see Sec.~\ref{subsec:rex}, i.e., in O-REX the quantity $\Delta_{ij}(X,X')$ is effectively substituted by the work $W_{ij}(\Gamma)$.

The O-REX move is still a swap of two configurations, as in REX; but the Metropolis--Hastings step lives in path space, as in out-of-equilibrium candidate Monte Carlo~\cite{Nilmeier_2011}. 
We then wish to find an acceptance probability $A_{\orex}$ for which the path-space detailed balance condition
\begin{equation}
\Pi_{ij}^{\sst{(0)}}(X_0)\,\mathcal{P}_\f(\Gamma)\,A_{\orex}(\Gamma)=\Pi_{ij}^{\sst{(\nstep)}}(X_{\nstep+1})\,\mathcal{P}_\r(\tilde{\Gamma})\,A_{\orex}(\tilde{\Gamma})
\end{equation}
is satisfied, like in Eq.~\eqref{eq:rex_detbal}. 
A subtle point comes from the fact that the proposal in REX was symmetric for an $i \leftrightarrow j$ exchange: this is true for O-REX only if $\mathcal{P}_\f(\Gamma)$ and $\mathcal{P}_\r(\tilde{\Gamma})$ are symmetric. This is ensured by using a protocol that is invariant under the combination of Monte Carlo time reversal plus the exchange of the two replicas. This explains our choice of protocol: performing $\nstep+1$ updates alternated by $\nstep$ changes in the couplings $\lambda$ makes the protocol palindromic, i.e., invariant under Monte Carlo time reversal.

Detailed balance requires $A_{\orex}(\Gamma)/A_{\orex}(\tilde{\Gamma})=\ee^{-W_{ij}(\Gamma)}$, which is satisfied by the Metropolis acceptance probability
\begin{equation}
    A_{\orex}(\Gamma)=\min\left\{1,\ee^{-W_{ij}(\Gamma)}\right\}.
    \label{eq:orex_acceptance}
\end{equation}
This is perfectly consistent with the REX case once $W_{ij}(\Gamma)\to\Delta_{ij}(X,X^\prime)$, see Eq.~\eqref{eq:rex_acceptance}.
If the move is accepted, the final configurations $x_i^{\sst{(\nstep+1)}}$ and $x_{j}^{\sst{(\nstep+1)}}$ are assigned to the parameters $\theta_j$ and $\theta_i$, respectively. Otherwise, both replicas are restored to their configurations before the move.

Let us now briefly discuss two interesting limits. For an instantaneous exchange, consisting of a single change of the joint action without updates, the work $W(\Gamma)$ reduces to $\Delta_{ij}(X,X^\prime)$ in Eq.~\eqref{eq:rex_delta}, and Eq.~\eqref{eq:orex_acceptance} recovers the standard REX acceptance probability in Eq.~\eqref{eq:rex_acceptance}. We thus label the REX case by $\nstep=0$. In the quasi-static limit, the system remains in equilibrium throughout the protocol, the dissipated work vanishes, $W(\Gamma)\to\Delta F=0$, and the acceptance probability approaches unity, $A_{\orex}(\Gamma)\to1$.

\subsection{Work and acceptance of O-REX moves}
\label{subsec:orex_scaling_theory}

The mean acceptance of O-REX moves depends on the distribution of the work. To make this dependence explicit, we apply Crooks' theorem to the joint two-replica system. We denote by $\rho_\f(W)$ and $\rho_\r(W)$ the probability distributions of the work induced by the forward and reverse joint path probabilities introduced in Sec.~\ref{subsec:orex} (we here drop the $ij$ subscripts to lighten the notation):
\begin{align}
    \rho_\f(W)&=\int \ddiff \Gamma\,\Pi^{(0)}(X_0)\,\mathcal{P}_\f(\Gamma)\,\delta\big(W-W(\Gamma)\big),
    \label{eq:work_distribution_f}\\
    \rho_\r(W)&=\int \ddiff\Gamma\,\Pi^{(\nstep)}(X_{\nstep+1})\,\mathcal{P}_\r(\tilde\Gamma)\,\delta\big(W-W(\tilde\Gamma)\big),
    \label{eq:work_distribution_r}
\end{align}
where $W(\tilde\Gamma)=-W(\Gamma)$. Inserting Eq.~\eqref{eq:orex_crooks} with $\Delta F=0$ into the definition of $\rho_\f(W)$, one obtains:
\begin{equation}
    \rho_\f(W)=\ee^{W}\rho_\r(-W).
    \label{eq:crooks_work}
\end{equation}
Since trajectories with $W<0$ are always accepted, while those with $W>0$ are accepted with probability $\ee^{-W}$, the mean acceptance can be split according to the sign of the work. Using Eq.~\eqref{eq:crooks_work} in the second integral:
\begin{equation}
    \begin{split}
    \langle A_{\orex}\rangle
    &=\int_{-\infty}^{0}\ddiff W\,\rho_\f(W)+\int_{0}^{\infty} \ddiff W\,\ee^{-W}\,\rho_\f(W)\\
    &=\mathrm{Prob}_\f(W<0)+\mathrm{Prob}_\r(W<0).
    \end{split}
    \label{eq:acceptance_crooks}
\end{equation}

For the symmetric protocol, the two terms in Eq.~\eqref{eq:acceptance_crooks} coincide: the reverse joint protocol is the forward one with the two replicas relabeled, so that $\rho_\r=\rho_\f$ and
\begin{equation}
    \langle A_{\orex}\rangle=2\,\mathrm{Prob}_\f(W<0).
    \label{eq:acceptance_2prob}
\end{equation}
The acceptance is then entirely determined by the probability of observing forward trajectories with negative dissipated work. Since $\Delta F=0$, these are the trajectories that locally violate the second law of thermodynamics, $\langle W\rangle_\f\geq\Delta F$, which holds only on average. Equation~\eqref{eq:acceptance_2prob} thus relates the acceptance of O-REX moves directly to the frequency of such violations.

To express this relation in a simpler form, we now assume that the distribution of $W$ is Gaussian, with mean $\mu=\langle W\rangle_\f$ and variance $\sigma_W^2$. This assumption is expected to be met close to the quasi-static regime, since the work of Eq.~\eqref{eq:orex_work} is a sum of many contributions accumulated along the protocol. Jarzynski's equality then fixes the variance in terms of the mean (recall that $\Delta F = 0$):
\begin{equation}
   \underbrace{\ee^{-\mu+\sigma_W^2/2}=}_{\substack{\text{Gaussianity}\\\text{assumption}}}  \left\langle \ee^{-W}\right\rangle_\f= \ee^{-\Delta F} = 1
    \quad\Longrightarrow\quad
    \sigma_W^2=2\mu = 2\langle W\rangle_\f .
    \label{eq:gaussian_jarzynski}
\end{equation}
Introducing the standard error function $\mathrm{erf}(z)= \frac{2}{\sqrt{\pi}} \int_{0}^{z} \ee^{-t^2} \, \ddiff t$, one gets,
\beq
\begin{aligned}
    \mathrm{Prob}_\f(W<0)
    =\frac{1}{2}\left\{1 -\mathrm{erf}\left(\frac{\mu}{\sqrt{2}\,\sigma_W}\right)\right\}
    &=\frac{1}{2}\left\{1-\mathrm{erf}\left(\frac{\sqrt{\mu}}{2}\right)\right\}\\
    & = \frac{1}{2}\left\{1-\mathrm{erf}\left(\frac{\sqrt{\langle W\rangle_\f}}{2}\right)\right\},
\end{aligned}
\eeq
which combined with Eq.~\eqref{eq:acceptance_2prob} gives
\begin{equation}
1-\langle A_{\orex} \rangle = \mathrm{erf}\left(\frac{\sqrt{\langle W\rangle_\f}}{2}\right).
    \label{eq:acceptance_erf}
\end{equation}
For small work, Eq.~\eqref{eq:acceptance_erf} gives
\begin{equation}
    1-\langle A_{\orex}\rangle\simeq\sqrt{\frac{\langle W\rangle_\f}{\pi}}, \qquad \qquad  \langle W\rangle_\f\ll1 .
    \label{eq:acceptance_limit_small}
\end{equation}
The same relations hold for standard REX, with $\langle W\rangle_\f$ replaced by $\langle\Delta\rangle$.

The O-REX acceptance is thus controlled by the average work, whose dependence on the parameters of the algorithm can be estimated as follows. Let us work assuming a linear dependence of $S_\theta$ on $\theta$:
\beq
S_\theta = S_0 + \theta O,
\eeq
with $O$ denoting the conjugate observable to the tempering parameter. In this case the work in
Eq.~\eqref{eq:orex_work} along the trajectory is:
\beq
W(\Gamma) = \frac{1}{\nstep}(\theta_j-\theta_i) \sum_{n=1}^{n_{\rm step}} \left[ O(x_{i}^{\sst{(n)}})-O(x_{j}^{\sst{(n)}}) \right]
\eeq
with $\Delta$ in
Eq.~\eqref{eq:rex_delta} being its instantaneous counterpart. For slow protocols, linear-response arguments~\cite{Sivak_2012} show that the average dissipated work is inversely proportional to $\nstep$ and quadratic in the variation of the control parameter:
\begin{equation}
  \langle W\rangle_\f \propto \frac{(\theta_j-\theta_i)^2}{\nstep}\,
  \mathrm{Var}(O),
  \label{eq:work_linear_response}
\end{equation}
where the proportionality constant depends on the relaxation time of $O$ under the intermediate dynamics.

The fluctuations of $O$ are extensive in the number $V_{\mathrm{dof}}$ of degrees of freedom coupled
to $\theta$, $\mathrm{Var}(O)\propto V_{\mathrm{dof}}$, while the spacing between adjacent parameters is fixed by the number of replicas, $\theta_j-\theta_i\sim 1/(\nrep-1)$.\footnote{Note that this relation also holds for ladders $\theta_i$ where the parameter spacing is not uniform, as the one we will use in this work.} Concerning the $1/\nstep$ behavior, we do not expect it to hold exactly in the range of values of
$\nstep$ considered here, since it requires the protocol to be much longer than the relaxation time of $O$ under the intermediate dynamics. With local updates, $O$ is in general coupled to long-wavelength
modes that relax diffusively, and which can produce a slowly decaying tail in its autocorrelation function and an effective exponent smaller than one at intermediate $\nstep$. We therefore parametrize the average work as follows:
\begin{equation}
  \langle W\rangle_\f = \frac{\langle\Delta\rangle}{\nstep^{\alpha}},
  \qquad \qquad
  \langle\Delta\rangle = k\,\frac{V_{\mathrm{dof}}}{(\nrep-1)^2},
  \label{eq:work_scaling}
\end{equation}
where $k>0$ and $0<\alpha\le 1$ can be determined through short numerical experiments, and $\langle\Delta\rangle$ is the action change for a standard REX swap proposal. A scaling of this form, with respect to both the size of the region affected by the transformation and the length of the protocol, has been observed for out-of-equilibrium transformations in $4d$ $\mathrm{SU}(3)$ gauge theory~\cite{Bulgarelli:2024brv,Bonanno:2025pdp}.

Combined with Eq.~\eqref{eq:acceptance_erf}, Eq.~\eqref{eq:work_scaling} predicts the
acceptance of any combination of $\nrep$, $\nstep$ and $V_{\mathrm{dof}}$ in terms of a single collective variable:
\begin{equation}
\langle A_{\orex}\rangle = 1- \mathrm{erf}\!\left(\frac{\sqrt{k}\,y}{2}\right) \simeq 1-\sqrt{\frac{k}{\pi}}\,y,  \qquad \qquad   y = \frac{1}{\nrep-1}\sqrt{\frac{V_{\mathrm{dof}}}{\nstep^{\alpha}}},
  \label{eq:acceptance_scaling}
    \end{equation}
once the parameters $k$ and $\alpha$ are fixed through short test runs. Recall that the approximation giving $\braket{A_{\orex}}\simeq 1 - \left(\sqrt{k/\pi}\right) y$ holds for $\sqrt{k}\,y\ll 1$, cf.~Eq.~\eqref{eq:acceptance_limit_small}.

\subsection{Cost model and optimal parameters}
\label{subsec:parameters}

Since the work spent along an O-REX protocol decreases with $\nstep$, this leads to a reduction in the number of replicas at fixed acceptance. Equations~\eqref{eq:work_scaling} and~\eqref{eq:acceptance_scaling} quantify this relation and, together with a cost model that we will discuss in the following, determine the optimal protocol length and the conditions for a cost gain. We write the computational cost as:
\begin{equation}
    \mathcal{C}_{\orex}=\nrep+2(\nrep-1)(\nstep+1)v,
    \label{eq:orex_cost}
\end{equation}
where the first term is the cost of standard REX with $\nrep$ replicas, and the second accounts for the additional updates along the $\nrep-1$ O-REX trajectories. Indeed, as explained before, each trajectory contains $\nstep+1$ updates per replica. The parameter $v$ is the cost of one protocol update relative to a full-lattice update (we work with $v\ll 1$, as we shall see later). For standard REX, since there is no out-of-equilibrium protocol, $\mathcal{C}_{\rex}=\nrep$.
In the case of O-REX, fixing a certain average target acceptance $\braket{A_{\orex}}=A^*$, Eq.~\eqref{eq:acceptance_scaling} fixes $y$, and hence the number of replicas:
\begin{equation}
    \nrep-1=\frac{1}{2\,\mathrm{erf}^{-1}(1-A^*)}\sqrt{\frac{k\,V_{\mathrm{dof}}}{\nstep^{\alpha}}} ,\qquad \nstep\ge1 .
    \label{eq:nrep_target}
\end{equation}
Inserted in Eq.~\eqref{eq:orex_cost}, this gives, up to a constant term:
\beq
\mathcal{C}\propto\nstep^{-\alpha/2}\,[1+2v(\nstep+1)].
\eeq
This scaling behavior has a clear interpretation: increasing $\nstep$ reduces the number of replicas needed, but makes each trajectory more expensive. The cost is minimized for $\nstep=\nstep^*$, where:
\begin{equation}
    \nstep^*=\frac{\alpha\,(1+2v)}{2v\,(2-\alpha)} .
    \label{eq:nstep_opt}
\end{equation}
Remarkably, the optimal length of the protocol depends only on $\alpha$ and $v$, and not on the target acceptance, on the prefactor $k$ or on $V_{\rm dof}$, which only fix the number of replicas through Eq.~\eqref{eq:nrep_target}. In particular, if $v$ is kept fixed, the same protocol length is optimal at all lattice spacings, as long as $\alpha$ does not change appreciably. Once $\nstep$ is fixed to $\nstep^*$, the number of replicas then follows from Eq.~\eqref{eq:nrep_target}.

Within the Gaussian approximation, REX and O-REX have the same swap acceptance when their mean work is equal. Under this condition, Eq.~\eqref{eq:work_scaling} allows one to establish:
\begin{equation}
    \nrep^{\sst\rm(OREX)}-1=( \nrep^{\sst\rm(REX)}-1)\nstep^{-\alpha/2}, \qquad \nstep \ge 1.
    \label{eq:orex_replica_matching}
\end{equation}
Inserting Eq.~\eqref{eq:orex_replica_matching} in Eq.~\eqref{eq:orex_cost} ultimately allows one to quantify the cost gain $G_{\mathrm{cost}}$ of O-REX in terms of REX:
\begin{equation}
    G_{\mathrm{cost}}\equiv\frac{\mathcal{C}_{\rex}}{\mathcal{C}_{\orex}}=\frac{\nrep^{\sst\rm(REX)}}{1+(\nrep^{\sst\rm(REX)}-1)\nstep^{-\alpha/2}\left[1+2v(\nstep+1)\right]}.
    \label{eq:orex_cost_gain}
\end{equation}
For $\nstep>1$ and $0<\alpha\le1$, the condition $G_{\mathrm{cost}}>1$ can be written as:
\begin{equation}
   0<v<\frac{\sqrt{\nstep}-1}{2(\nstep+1)} \le \frac{\sqrt{2}-1}{4}\approx 0.104,
    \label{eq:orex_v_bound}
\end{equation}
and
\begin{equation}
    \alpha_{\min}(v,\nstep)\equiv\frac{2\log\left[1+2v(\nstep+1)\right]}{\log(\nstep)}<\alpha\le1.
    \label{eq:orex_alpha_bound}
\end{equation}
The relative protocol update cost $v$ must therefore be sufficiently cheap (10\% of the full update at most), and the work must decrease sufficiently rapidly with $\nstep$, to make the reduction in replica number worthwhile. This estimate compares computational costs at fixed acceptance but does not account for changes in autocorrelation times. We therefore also assess the overall efficiency gain numerically using the following figure of merit:
\beq
\mathcal{C}\,\tauint(Q^2).
\eeq
This product also takes into account the autocorrelation of the slowest modes of our simulations, i.e., the topological ones.

\subsection{Application to PTBC}
\label{subsec:orex_ptbc}

In this work, we apply the O-REX framework to Parallel Tempering on Boundary Conditions (PTBC) to define an Out-of-equilibrium PTBC (O-PTBC). Here the tempering parameter $\theta$ is mapped to the defect boundary condition parameter $c$, $S_\theta$ is mapped to the replica action $S_{\L}^{\sst{(c)}}$ in Eq.~(2.7), and a configuration $x$ to the fields $(z,\bar z,U)$.

A ladder of $\nrep$ replicas with $1=c_0>c_1>\cdots>c_{\nrep-1} = 0$ interpolates between the physical system with \PBC{} ($c_0=1$) and the system with an open defect ($c_{\nrep-1}=0$). Exchanges are always proposed between replicas with adjacent couplings $c_i$ and $c_{i+1}$. As explained earlier, the idea is that physical observables are measured only on the periodic replica, $c_0=1$, free of boundary effects, while the remaining replicas let topological excitations enter and leave through the defect. To let the defect explore the whole lattice, after each round of exchanges its position is effectively moved by one site along a randomly chosen positive direction. In practice, this is implemented by translating the configuration of the periodic replica by one lattice spacing~\cite{Hasenbusch:2017unr}. Since the action is linear in $c$, see Eq.~\eqref{eq:cpn_action_linear_c}, the interpolating action of Eq.~\eqref{eq:orex_action} corresponds to driving the two defect couplings linearly in opposite directions:
\begin{equation}
    c_i^{(n)}=(1-\lambda_n)\,c_i+\lambda_n\,c_j,
    \qquad
    c_j^{(n)}=(1-\lambda_n)\,c_j+\lambda_n\,c_i.
\end{equation}

Concerning the standard PTBC, one simply proposes and accepts a swap of two replicas $(i,j=i+1)$ with probability:
\beq
P_{\sst\rm swap}^{\sst \rm (PTBC)} &=& \min\left\{1, \ee^{-\Delta_{ij}}\right\},\\
\Delta_{ij} &=& -(c_i-c_j)\left\{S_{\L}^{\sst\rm (\mathcal{D})}(x_i)-S_{\L}^{\sst\rm (\mathcal{D})}(x_j)\right\}.
\eeq
For O-PTBC, instead, the swap acceptance is modified as:
\beq
P_{\sst\rm swap}^{\sst \rm (OPTBC)} &=& \min\left\{1, \ee^{-W_{ij}(\Gamma)}\right\},
\eeq
with the work given by
\begin{equation}
    W_{ij}(\Gamma)=\frac{c_j-c_i}{\nstep}\sum_{n=1}^{\nstep}\left[S_{\L}^{\sst{(\mathcal{D})}}(x_{i}^{\sst (n)})-S_{\L}^{\sst{(\mathcal{D})}}(x_{j}^{\sst (n)})\right].
    \label{eq:orex_work_ptbc}
\end{equation}
In both cases, the swap probability is fully determined by $S_{\L}^{\sst{(\mathcal{D})}}$, i.e., by only the terms in the action containing a link crossing the defect, cf.~Eq.~\eqref{eq:cpn_action_linear_c}. The number of degrees of freedom coupled to $c$ is therefore proportional to the defect size, and in the following we set $V_{\mathrm{dof}}=V_{\dd}\equiv L_\dd$.

\section{Numerical results}\label{sec:results}

In this section, we will present numerical results comparing the performance and computational cost of O-REX with those of the standard REX in the $N=21$ model. We will compare how the costs of the two algorithms scale towards the continuum limit, and the continuum limits of the topological susceptibility.

\subsection{Simulation details}
\label{subsec:sim_details}

We consider five different simulation points drawn from a line of constant physics with $\ell/\xi = aL/\xi\simeq21$ and $\ell_\dd/\xi = a L_\dd/\xi \simeq 1.3$, corresponding to five different lattice spacings ranging from $\xi/a\sim 4$ up to $\xi/a \sim 7$. The volume was chosen according to the results of~\cite{Berni:2019bch}, where it was shown that results obtained for $\ell/\xi \gtrsim 10$ are all statistically consistent with each other at the per-mille precision. The defect size was chosen following the results of Ref.~\cite{Bonanno:2022hmz}, where it was shown that $\ell_\dd\sim\mathcal{O}(\xi)$ is sufficiently large to enable an ergodic sampling of the topological charge. Concerning the calculation of the topological susceptibility, we fixed the smoothing radius as $n_{\rm cool} \times (a/\xi)^2 \simeq 1$, where $n_{\rm cool} \times (a/\xi)^2 \propto \Rs^2/\xi^2$. All simulation parameters are listed in Tab.~\ref{tab:physical}.

\begin{table}[!t]
\centering
\begin{tabular}{cccccccc}
\hline
$L$ & $\beta$ & $\xi/a$ & $\ell/\xi$ & $L_\dd$ & $\ell_\dd/\xi$ & $L_\R$ & $\ell_\R/\xi$ \\
\hline
88  & 0.660 & 4.2129(78) & 20.9 & 5 & 1.19 & 8 & 1.90 \\
102 & 0.680 & 4.7561(99) & 21.4 & 6 & 1.26 & 9 & 1.89 \\
114 & 0.700 & 5.422(15)  & 21.0 & 7 & 1.29 & 10 & 1.84 \\
128 & 0.718 & 6.030(13)  & 21.2 & 8 & 1.33 & 11 & 1.82 \\
150 & 0.741 & 6.979(11)  & 21.5 & 9 & 1.29 & 13 & 1.86 \\
\hline
\end{tabular}
\caption{Simulation parameters along the line of constant physics with fixed $\ell/\xi$ and $\ell_\dd/\xi$ for the $N=21$ model. Results for the correlation length $\xi/a$ come from Ref.~\cite{Bonanno:2022hmz}, $L_\dd$ is the defect length in lattice units, while $L_\R$ is the half-size (in lattice units) of the lattice patch that is updated along out-of-equilibrium O-REX trajectories, see Eq.~\eqref{eq:patch_volume}. These choices correspond to fixed $v\equiv V_{\rm patch}/V\simeq 0.045$, see the text below.}
\label{tab:physical}
\end{table}

Let us now discuss how we practically implemented the two algorithmic schemes we want to compare. All replicas are evolved independently at equilibrium with local updates. One Monte Carlo updating step consists of: (i) one updating step of every replica, made of four over-relaxation (OR) sweeps followed by one heat-bath (HB) sweep over the whole lattice, for both the scalar fields $z(x)$ and the gauge links $U_\mu(x)$ (each sweep is done sequentially in lexicographic order); (ii) one round of REX/O-REX moves, proposed sequentially on the pairs of adjacent replicas $(c_0,c_1),(c_1,c_2),\ldots,(c_{\nrep-2},c_{\nrep-1})$; (iii) the translation in a random direction by one lattice spacing of the periodic replica. All observables are measured on the periodic replica ($c_0=1$) every 10 updating steps for both REX and O-REX.

The defect couplings $c_r$ of the replica ladder are not equally spaced, but have been determined according to the optimal-schedule procedure of Ref.~\cite{Syed:2024}, which equalizes the swap acceptance rates along the ladder. The optimal couplings $c_r$ are determined in standard REX simulations and reused, unchanged, for O-REX. More details on the determination of the optimal schedule for the defect coupling parameters $c_r$ are given in Appendix~\ref{app:optimal_protocol}.

Motivated by the bound in Eq.~\eqref{eq:orex_v_bound}, the updates along the out-of-equilibrium trajectory are not performed on the whole lattice, but only on a rectangular patch centered on the defect, containing $2L_\R+1$ sites in the spatial direction and $2L_\R+L_\dd$ sites in the temporal one, i.e.\ extending by $L_\R$ sites beyond the defect on each side. Its volume is therefore:
\begin{equation}
    a^2V_{\mathrm{patch}}=(2L_\R+1)\,(2L_\R+L_\dd).
    \label{eq:patch_volume}
\end{equation}
Since only the defect couplings change along the protocol, the system is driven out of equilibrium mainly close to the defect, and restricting the updates to this region substantially reduces the cost of each trajectory. This is in the same spirit as the defect coupling layers of Refs.~\cite{Bulgarelli:2024yrz,Bonanno:2025pdp}. We choose $\ell_{\R}=a L_\R\simeq2\xi$, corresponding to $v=V_{\mathrm{patch}}/V\simeq0.045$, motivated by stochastic locality~\cite{Luscher:2017cjh}: the effect of a change of the defect couplings decays exponentially over a distance of order $\xi$. We verified the validity of this choice \emph{a posteriori}: the check is shown in Appendix~\ref{app:local_updates}, where we compare the O-REX performance at different patch sizes for $L=114$ and $L=150$. Clearly, larger patches reduce $\tauint(Q^2)$ in the periodic replica, but the overhead cost of the protocol also grows with $V_{\mathrm{patch}}$. In the end, the figure of merit $\mathcal{C}\,\tauint(Q^2)$ is lowest for $\ell_\R\simeq2\xi$.

An out-of-equilibrium protocol consists of $\nstep+1$ patch updates, alternated with $\nstep$ changes of the defect coupling parameters, which are interpolated linearly. Each patch update consists of four OR sweeps and one HB sweep restricted to the patch. We choose this sequence or its reverse with probability $1/2$. The reverse is obtained by reversing the order of all elementary updates (i.e., sweeping sites sequentially in reverse lexicographic order). We recall that the reason for this peculiar choice, as well as for the unequal number of patch updates ($\nstep+1$) and defect coupling parameter evolutions ($\nstep$) is to make the full protocol symmetric, see Sec.~\ref{subsec:orex}. Notice that the restricted updates satisfy detailed balance too, thus O-REX remains exact.

For each of the simulation parameters in Tab.~\ref{tab:physical}, several simulations were performed to explore different REX/O-REX setups. For each simulation, $\mathcal{O}(100)$ independent REX/O-REX streams have been run. For each one, we started from a cold configuration and discarded the first $10^4$ Monte Carlo steps for thermalization. The complete list of simulations is given in Appendix~\ref{app:runs}.

\subsection{Tests of the work-acceptance relations}
\label{subsec:orex_scaling}

\begin{figure}[!t]
\centering
\includegraphics[width=0.85\textwidth]{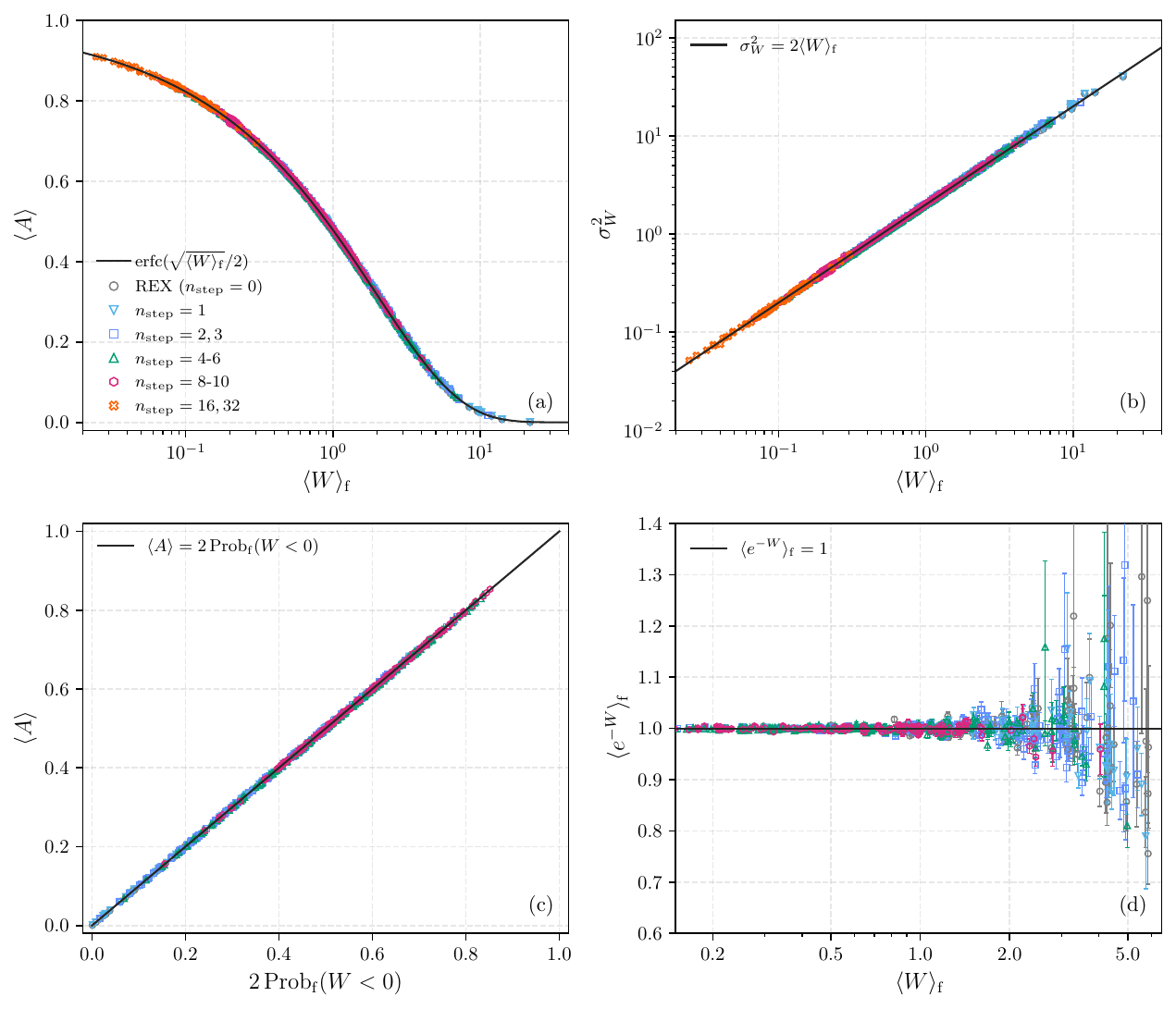}
\caption{Top left: acceptance against average work, cf.~Eq.~\eqref{eq:acceptance_erf}. Top right: variance of the work against its mean, compared with the prediction $\sigma_W^2=2\langle W\rangle_\f$ in Eq.~\eqref{eq:gaussian_jarzynski}.
Bottom left: acceptance against twice the probability of a negative work, cf.~Eq.~\eqref{eq:acceptance_2prob}. Bottom right: check of Jarzynski's equality, Eq.~\eqref{eq:jarzynski} with $\Delta F=0$. Data points are obtained from several simulations with $L=114$ ($L_\R$ ranging from 1 to $L/2$, $\nrep$ from 5 to 14, both linear and optimal ladders) and $L=150$
($L_\R=13,14,50$, $\nrep=7$, optimal ladders).}
\label{fig:acceptance}
\end{figure}

In this section, we test the relations between the work and the O-REX swap acceptance rate derived in Sec.~\ref{subsec:orex_scaling_theory}. In Fig.~\ref{fig:acceptance} we show several measurements for $\mathcal{O}(2500)$ pairs of adjacent replicas across $\mathcal{O}(300)$ simulation setups, with $0\le\nstep\le32$, $L=114$ and $150$, and patches ranging from $L_\R=1$ to $L_\R=L/2$:
\begin{itemize}
\item The top left panel shows the mean acceptance $\langle A{_\orex}\rangle$ as a function of the mean work $\langle W\rangle_{\f}$. The data follow the Gaussian prediction of Eq.~\eqref{eq:acceptance_erf} over the range $0.02\lesssim\langle W\rangle_\f\lesssim20$, with a root-mean-square (rms) deviation of 0.19 percentage points.

\item The top right panel compares the work variance with its mean, testing the Gaussian relation $\sigma_W^2=2\langle W\rangle_\f$ of Eq.~\eqref{eq:gaussian_jarzynski}. The ratio $\sigma_W^2/(2\langle W\rangle_\f)$ has a median of 1.007, with the 5th and 95th percentiles at 0.98 and 1.04.

\item The bottom left panel compares $\langle A_{\orex}\rangle$ with $2\,\mathrm{Prob}_\f(W<0)$. The exact relation of Eq.~\eqref{eq:acceptance_2prob}, which does not require a Gaussian work distribution, is satisfied with an rms deviation of 0.15 percentage points.

\item The bottom right panel shows $\langle \ee^{-W}\rangle_\f$ as a function of $\langle W\rangle_\f$. The values agree with unity, as required by Jarzynski's equality~\eqref{eq:jarzynski}, with an rms deviation of 1.1 standard deviations. Further tests of the Gaussianity of the work distributions and of Crooks' relation are presented in Appendix~\ref{app:work_distributions}.
\end{itemize}

\begin{figure}[!t]
    \centering
    \includegraphics[width=\textwidth]{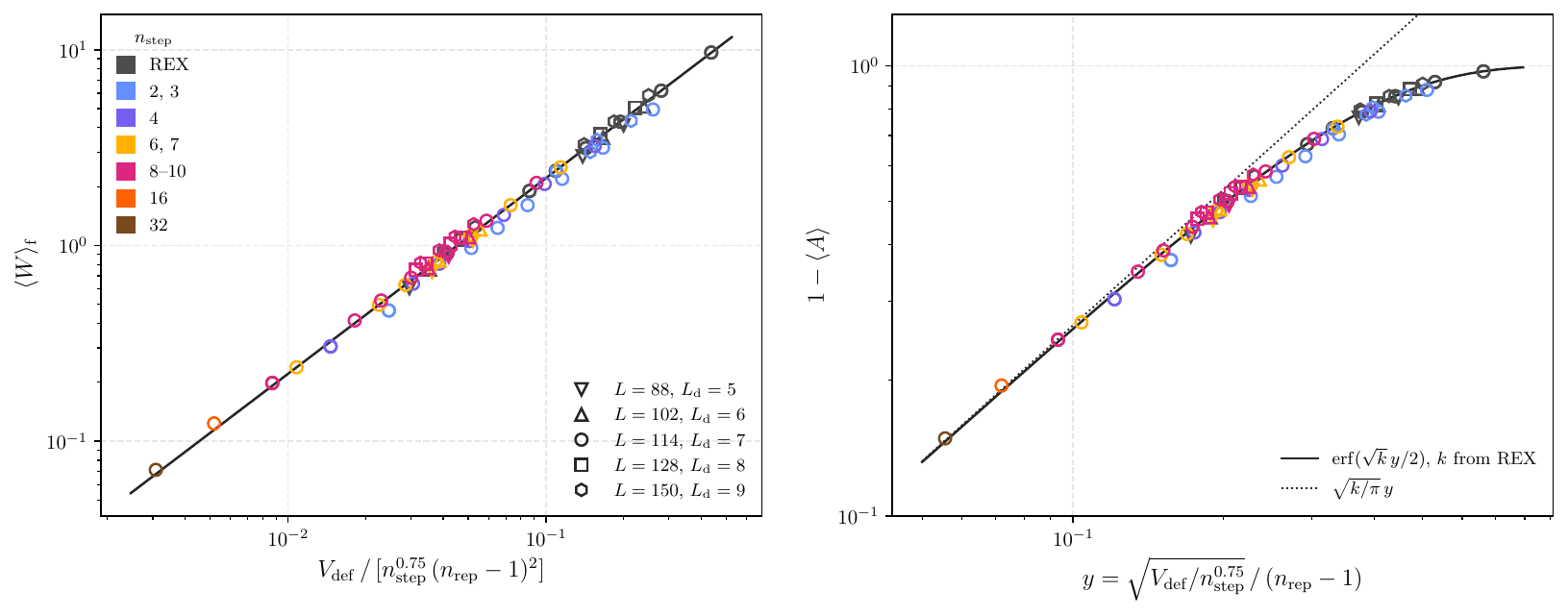}
    \caption{Left: average O-REX work against the predicted scaling variable $y^2 = V_{\mathrm{def}}/[\nstep^{0.75}(\nrep-1)^2]$, for the five lattice spacings of Tab.~\ref{tab:physical}. Different colors denote different protocol lengths, and standard REX ($\nstep=0$) is placed at $\nstep=1$, where O-REX has the same average work. The line is Eq.~\eqref{eq:work_scaling} with $\alpha=0.75$ and $k=22.1$ (obtained by fitting the REX data points). Right: swap rejection rate against $y$, cf.~Eq.~\eqref{eq:acceptance_scaling}. The solid line is the prediction $\mathrm{erf}(\sqrt{k}\,y/2)$, the dotted one its small-$y$ limit $\sqrt{k/\pi}\,y$.}
    \label{fig:scaling}
\end{figure}

Fig.~\ref{fig:scaling} tests Eq.~\eqref{eq:work_scaling} on 82 setups at the five lattice spacings of Tab.~\ref{tab:physical}, with different numbers of replicas, protocol lengths and defect sizes. A fit to the REX points gives $k=22.1(2)$. With $k$ fixed to this value, a fit to the O-REX points gives $\alpha=0.75(2)$, where the error covers the variation with the range of $\nstep$ included in the fit. With $k=22.1$ and $\alpha=0.75$, the data follow the predicted scaling over a range of values for the scaling variable $y=\sqrt{V_{\mathrm{def}}/[\nstep^{0.75}(\nrep-1)^2]}$ spanning two orders of magnitude in the mean work. The REX points are plotted at $\nstep=1$, where Eq.~\eqref{eq:work_scaling} gives $\langle W\rangle_\f=\langle\Delta\rangle$. The value $\alpha=0.75$ is an effective exponent for the finite protocol lengths considered here, below the linear-response value $\alpha=1$, as discussed in Sec.~\ref{subsec:orex_scaling_theory}. The right panel shows the rejection probability $1-\langle A_{\orex}\rangle$ against the scaling variable $y$. The data follow Eq.~\eqref{eq:acceptance_scaling} with the same values of $k$ and $\alpha$, without additional parameters. 

\subsection{Parameter optimization, efficiency and scaling with lattice size}
\label{subsec:orex_cost}

We now use the work scaling tested above to choose the number of replicas and the protocol length, with the patch size described in Sec.~\ref{subsec:sim_details}. Since $v=V_{\mathrm{patch}}/V$, the bound in Eq.~\eqref{eq:orex_v_bound} limits the patch volume for which a cost gain is possible at a given protocol length. The study of the O-REX performance as a function of $v$ is reported in Appendix~\ref{app:local_updates}. To include the effect on autocorrelation times, we compare different setups using the figure of merit $\mathcal{C}\,\tauint(Q^2)$. Autocorrelation times are quoted in measurement intervals of ten Monte Carlo steps, and we omit this common factor of ten from the cost measure. As anticipated in Sec.~\ref{subsec:sim_details}, we found that the best performances are obtained for $\ell_\R \simeq 2 \xi$. This corresponds to $v\simeq 0.045$, below the bound $v \lesssim 0.1$.

\begin{figure}[!t]
    \centering
    \includegraphics[width=\textwidth]{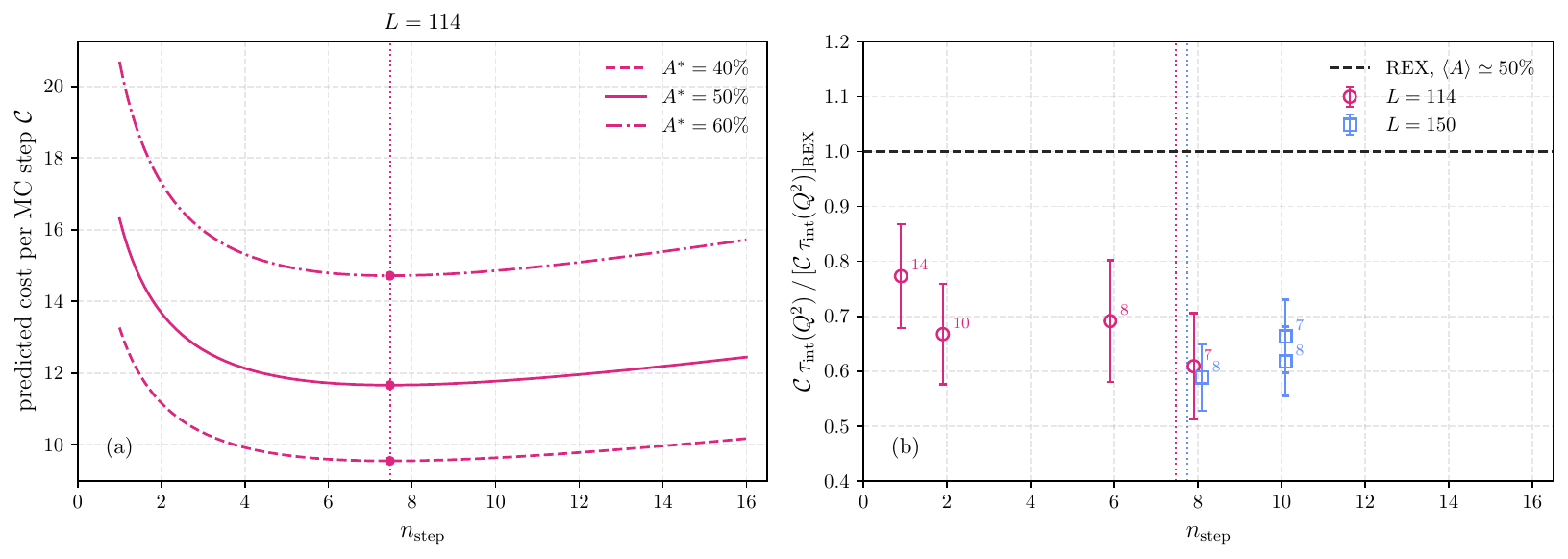}
    \caption{Left: cost $\mathcal{C}$ at $L=114$ predicted by Eqs.~\eqref{eq:orex_cost} and~\eqref{eq:nrep_target}, for target acceptances of 40\%, 50\% and 60\%. The dots mark the minima and the vertical line is $\nstep^*$, cf.~Eq.~\eqref{eq:nstep_opt}. Right: measured $\mathcal{C}\,\tauint(Q^2)$ of O-REX setups with acceptance within 2.5 ($L=114$, $L_\R=10$) or 4 ($L=150$, $L_\R=13$) percentage points of 50\%, relative to standard REX with acceptance close to $50\%$ ($\nrep=14$ and $16$) at the same lattice spacings. Text labels report the number of replicas $\nrep$, and the vertical lines mark $\nstep^*$ at the two lattice spacings. At $L=150$, $\nstep=8$ and $\nstep=10$ correspond to the fixed and the rescaled protocol lengths, respectively.}
    \label{fig:optimal}
\end{figure}

After having fixed $v$, Eq.~\eqref{eq:nstep_opt} determines the optimal protocol length, while the number of replicas depends also on the chosen target acceptance. We choose $A^*=50\%$, guided by the comparison in Appendix~\ref{app:acceptances}: at $L=114$, the lowest computational costs are found between 50\% and 60\% and are compatible within errors. Using the fitted values $k=22.1$ and $\alpha=0.75$, Eqs.~\eqref{eq:nstep_opt} and~\eqref{eq:nrep_target} give $\nstep^*\simeq7.5$ and $\nrep^*\simeq7.1$ at $L=114$, i.e.\ $\nstep=8$ and $\nrep=7$, to be compared with $\nrep=14$ required with the standard PTBC to get the same acceptance. Remarkably, the required number of replicas is halved. Note also that the optimal value of $\nstep$ is not sensitive to the precise value of $\alpha$. For example, $\alpha=0.65$ would give $\nstep^*\simeq6$, and the new minimum cost would change by just 1\%. Thus, the optimization of the algorithm is fairly robust under changes of the work scaling.

To test how the parameters determined at $L=114$ transfer to a finer lattice, we extend the comparison to $L=150$. Along the line of constant physics, the defect length in lattice units increases from $L_\dd=7$ to 9, and we choose $L_\R=13$ to keep $v$ approximately unchanged. At fixed acceptance, Eq.~\eqref{eq:nrep_target} requires $(\nrep-1)^2\nstep^\alpha\propto L_\dd$, i.e., a larger defect can be accommodated by increasing the replica number, the protocol length, or both. For fixed $v$, Eq.~\eqref{eq:nstep_opt} predicts an approximately constant optimal protocol length, with $\nrep-1$ increasing as $\sqrt{L_\dd}$. The predicted values are $\nstep^*\simeq7.7$ and $\nrep^*\simeq 7.9$ at 50\% acceptance, again suggesting $\nstep=8$ and $\nrep=8$. We also test $\nstep=10$, obtained by increasing the reference protocol length proportionally to $L_\dd$, to check whether the additional updates improve the overall efficiency. Both protocol lengths are tested with $\nrep=7$ and 8. With standard PTBC, at the $L=150$ point one again needs double the replicas, $\nrep=16$, to get the same target acceptance.

The left panel of Fig.~\ref{fig:optimal} shows the predicted cost at $L=114$. Its minimum is independent of the target acceptance and is shallow: the cost remains within 1\% of the minimum for $6\lesssim\nstep\lesssim10$. Note that for $\nstep=8$ both conditions for a cost gain of Sec.~\ref{subsec:parameters} are satisfied: Eq.~\eqref{eq:orex_v_bound} requires $v\lesssim 0.10$, and our patch has $v\simeq0.045$; Eq.~\eqref{eq:orex_alpha_bound} gives $\alpha_{\min}\simeq0.56$, below the measured $\alpha=0.75$. With these parameters Eq.~\eqref{eq:orex_cost_gain} predicts $G_{\mathrm{cost}}\simeq1.2$.

The right panel of Fig.~\ref{fig:optimal} shows the figure of merit $\mathcal{C}\,\tauint(Q^2)$ relative to standard REX for setups with acceptance close to $50\%$. The lowest measured values occur at $(\nrep,\nstep)=(7,8)$ for $L=114$ and $(8,8)$ for $L=150$, in agreement with our predictions. At $L=114$, the best setup gives $\mathcal{C}\,\tauint(Q^2)=103(14)$, compared with $169(15)$ for standard REX with 14 replicas at the same acceptance, corresponding to a gain of $1.6(3)$, or roughly 40\% lower computational cost at fixed statistical precision. At $L=150$, the corresponding values are $172(14)$ and $287(15)$, giving a similar gain of $1.67(16)$. Using 16 REX replicas to match the acceptance at $L=150$ gives a gain of about $1.7$. On this lattice, the results with $\nstep=8$ and $10$ are compatible within errors, so the longer protocol shows no clear efficiency advantage.

For the best setup at $L=114$, the cost at $L=150$ decreases from $\mathcal{C}_{\rex}=16$ to $\mathcal{C}_{\orex}\simeq13.3$, giving $G_{\mathrm{cost}}\simeq1.20$, consistent with Eq.~\eqref{eq:orex_cost_gain}. The autocorrelation time decreases from $18.3(11)$ to $12.9(11)$, providing a further gain of about $1.4$. These factors multiply to give the overall efficiency gain. Finally, for both values of $L$ the number of replicas is halved.

Even at $\nstep=1$, where the replica number and the swap acceptance are unchanged, the cost measure at $L=114$ is about 23\% lower than for standard REX. The additional patch updates therefore improve decorrelation enough to compensate for their cost. This also shows that $G_{\mathrm{cost}}>1$ is not necessary for an overall efficiency gain. Nonetheless, the optimal protocol length predicted from the minimization of the cost agrees with the measured optimum within the precision of our data.

\subsection{Scaling toward the continuum limit}
\label{subsec:continuum}

\begin{table}[!t]
\centering
\footnotesize
\begin{tabular}{ccccccccc}
\hline
$L$ & $\xi/a$ & & $\nrep$ & $\nstep$ & $\langle A\rangle$ (\%) & $\tauint(Q^2)$ & $\mathcal{C}\,\tauint(Q^2)$ & $\xi^2\chi\times10^3$ \\
\hline
\multirow{4}{*}{88}  & \multirow{4}{*}{4.2129(78)} & MCMC  & 1     & -- & --        & 396(10)   & 396(10)  & -- \\
                     &                             & REX   & 14    & 0  & 58.2      & 6.65(63)  & 93.1(8.8) & 7.785(48) \\
                     &                             & O-REX & 6     & 8  & 51.1      & 5.65(75)  & 57.3(7.6) & \multirow{2}{*}{7.770(43)} \\
                     &                             & O-REX & 7     & 6  & 54.7      & 5.78(60)  & 62.8(6.5) & \\
\hline
\multirow{4}{*}{102} & \multirow{4}{*}{4.7561(99)} & MCMC  & 1     & -- & --        & 1270(45)  & 1270(45) & -- \\
                     &                             & REX   & 14    & 0  & 54.0      & 10.65(72) & 149(10)  & 7.666(47) \\
                     &                             & O-REX & 7     & 8  & 53.9      & 6.78(82)  & 79.5(9.6) & \multirow{2}{*}{7.684(41)} \\
                     &                             & O-REX & 7     & 7  & 52.1      & 9.6(1.4)   & 108(16)  & \\
\hline
\multirow{3}{*}{114} & \multirow{3}{*}{5.422(15)}  & MCMC  & 1     & -- & --        & 3760(210) & 3760(210) & -- \\
                     &                             & REX   & 14    & 0  & 50.1      & 12.1(1.1)  & 169(15)  & 7.774(53) \\
                     &                             & O-REX & 7     & 8  & 49.9      & 8.8(1.2)   & 103(14)  & 7.797(49) \\
\hline
\multirow{4}{*}{128} & \multirow{4}{*}{6.030(13)}  & MCMC  & 1     & -- & --        & 10240(730) & 10240(730) & -- \\
                     &                             & REX   & 14    & 0  & 46.5      & 17.1(1.2)  & 239(17)  & 7.739(51) \\
                     &                             & O-REX & 7     & 8  & 46.2      & 12.1(2.0)  & 140(23)  & \multirow{2}{*}{7.716(40)} \\
                     &                             & O-REX & 7     & 9  & 47.9      & 11.3(1.2)  & 136(14)  & \\
\hline
\multirow{4}{*}{150} & \multirow{4}{*}{6.979(11)}  & MCMC  & 1     & -- & --        & --        & --        & -- \\
                     &                             & REX   & 16    & 0  & 49.5      & 18.3(1.1)  & 292(18)  & 7.739(38) \\
                     &                             & O-REX & 8     & 8  & 49.6      & 12.9(1.1)  & 172(14)  & \multirow{2}{*}{7.765(31)} \\
                     &                             & O-REX & 7     & 10 & 46.1      & 15.4(1.2)  & 194(15)  & \\
\hline
\end{tabular}
\caption{Simulations along the line of constant physics used in Figs.~\ref{fig:continuum}, \ref{fig:continuum_gain} and~\ref{fig:chi_continuum}. For REX, the runs with acceptance closest to $50\%$; for O-REX, $\nstep=8$ and $\nstep\simeq8\,L_\dd/7$ (at $L=114$ the two coincide). For standard MCMC, $\mathcal{C}=1$; at $L=150$ a reliable estimate of $\tauint(Q^2)$ was computationally too expensive. For REX and OREX, $\xi^2\chi$ has been averaged over the available runs at each lattice spacing. Units for $\tauint(Q^2)$ are the same as in Tab.~\ref{tab:runs}.}
\label{tab:continuum}
\end{table}

We now extend the comparison to all five lattice spacings in Tab.~\ref{tab:physical}, following the parameter choice described in Sec.~\ref{subsec:orex_cost}. We keep $v = V_{\mathrm{patch}}/V$ approximately fixed and consider two protocol lengths: $\nstep=8$ and $\nstep\simeq8\,L_\dd/7$, rounded to the nearest integer. For O-REX at fixed $\nstep$, the number of replicas is scaled as $\nrep\propto\sqrt{L_\dd}$~\cite{Hukushima:1995mcr}, while for $\nstep\simeq8\,L_\dd/7$ Eq.~\eqref{eq:nrep_target} gives $\nrep=7$ at all lattice spacings. For REX, we use the runs with acceptance closest to $50\%$, i.e.\ $\nrep=14$ at the four coarsest lattice spacings and $\nrep=16$ at the finest one. All simulations are listed in Tab.~\ref{tab:continuum}.

\begin{figure}[!t]
    \centering
    \includegraphics[width=0.75\textwidth]{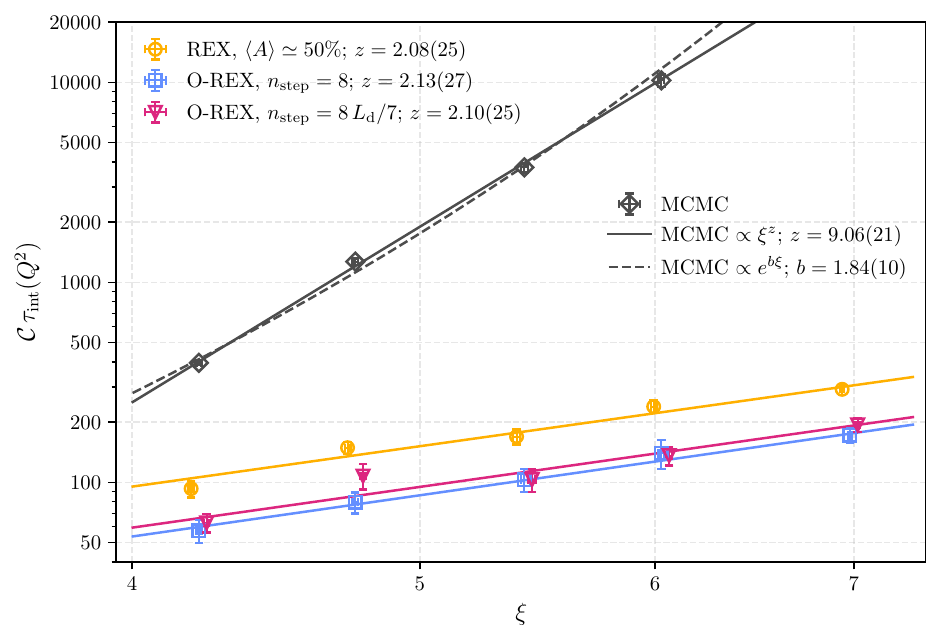}
    \caption{$\mathcal{C}\,\tauint(Q^2)$ as a function of the correlation length, for standard REX, O-REX with $\nstep=8$ or $\nstep\simeq8\,L_\dd/7$, and standard MCMC on a single periodic replica (four coarsest lattice spacings). Both REX and O-REX results correspond to an acceptance close to 50\%. Lines are fits to $A\xi^z$; for MCMC, the dashed line is a fit to $Be^{b\xi}$.}
    \label{fig:continuum}
\end{figure}

\begin{figure}[!t]
    \centering
    \includegraphics[width=0.7\textwidth]{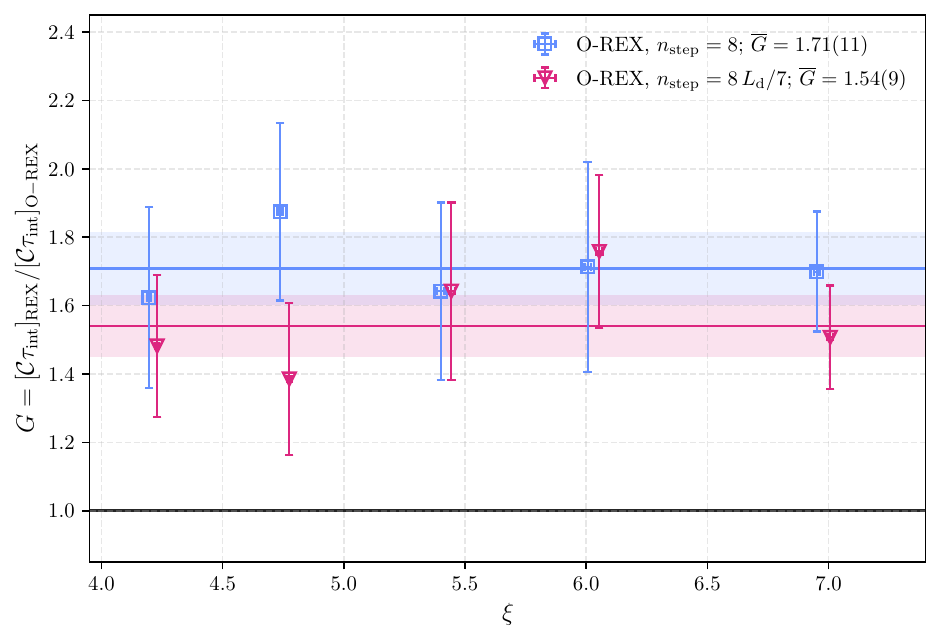}
    \caption{Gain of O-REX relative to REX; colored lines and bands show the weighted means and their $1\sigma$ uncertainties. Points are slightly displaced horizontally for readability.}
    \label{fig:continuum_gain}
\end{figure}

Fig.~\ref{fig:continuum} shows $\mathcal{C}\,\tauint(Q^2)$ as a function of the correlation length for the two prescriptions discussed in Sec.~\ref{subsec:orex_cost}. Fits to $\mathcal{C}\,\tauint(Q^2)\propto\xi^z$ give $z=2.08(25)$ for REX, $z=2.13(27)$ for O-REX with fixed protocol length, and $z=2.10(25)$ for the protocol length scaled with $L_\dd$. The exponents are compatible within errors. For comparison, Fig.~\ref{fig:continuum} also shows standard MCMC, i.e.\ a single replica with periodic boundary conditions and local updates, for which $\mathcal{C}=1$. Its cost grows much faster: at the finest lattice spacing $\xi/a\sim 7$, we were not even able to provide a reliable estimate of $\tauint$ for $Q^2$. With available data, a power-law fit gives an exponent $z=9.06(21)$, with a reduced chi-squared of $\tilde{\chi}^2/\mathrm{dof}=1.7$, while an exponential fit, $\mathcal{C}\,\tauint(Q^2)\propto e^{b\xi}$, gives $b=1.84(10)$, with a reduced chi-squared of $\tilde{\chi}^2/\mathrm{dof}=9.4$. At the next-to-finest lattice spacing $\xi/a\simeq6$, standard MCMC is about 43 times more expensive than REX and 73 times more expensive than O-REX with $\nstep=8$. Fig.~\ref{fig:continuum_gain} shows that the gain is consistent with a constant over the range studied, with weighted averages $\bar{G}=1.71(11)$ and $1.54(9)$ for the two O-REX prescriptions, respectively.

\begin{figure}[!t]
    \centering
    \includegraphics[width=0.65\textwidth]{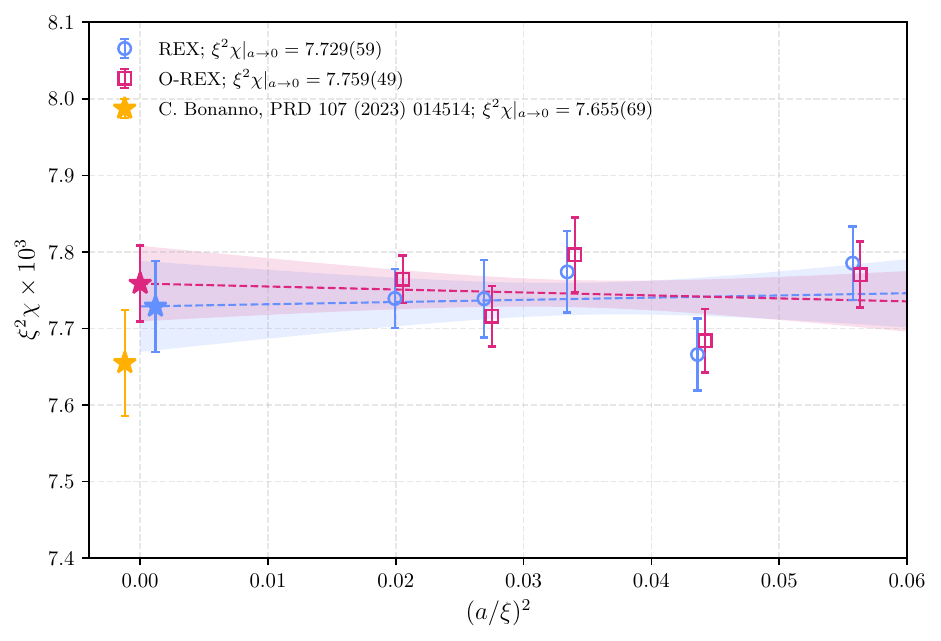}
    \caption{Continuum limit of $\xi^2\chi$ from the REX (circles) and O-REX (squares) simulations of this work, averaged separately at each lattice spacing. Lines and bands are linear fits in $(a/\xi)^2$ with their $1\sigma$ uncertainty; the stars are the continuum values, together with that of Ref.~\cite{Bonanno:2022hmz}, slightly displaced horizontally for readability.}
    \label{fig:chi_continuum}
\end{figure}

We conclude this section by studying the continuum limit of the topological susceptibility. Fig.~\ref{fig:chi_continuum} shows the continuum limit extrapolation of $\xi^2\chi$ from our REX and O-REX simulations, together with the continuum determination of Ref.~\cite{Bonanno:2022hmz}. Linear extrapolations in $(a/\xi)^2$ give $\xi^2\chi=7.729(59)\times10^{-3}$ for REX and $7.759(49)\times10^{-3}$ for O-REX, both with a slope compatible with zero (signaling the presence of mild lattice artifacts). The two continuum limits agree with each other, and are within $0.8\sigma$ and $1.2\sigma$, respectively, of the continuum value $7.655(69)\times10^{-3}$ obtained in~\cite{Bonanno:2022hmz}.

\section{Conclusions}
\label{sec:conclusions}

In this work we discussed the Out-of-equilibrium Replica Exchange (O-REX) sampling strategy, and we applied it to Parallel Tempering on Boundary Conditions (PTBC). The new algorithm was tested for the $2d$ $\CP^{N-1}$ model with $N=21$. Instead of swapping two configurations instantaneously, O-REX exchanges their defect couplings gradually along a short protocol, during which the fields around the defect are updated. Since the move is accepted with a Metropolis probability depending on the work done along the protocol, the algorithm is exact for any protocol length, as we verified directly against standard Replica Exchange (REX) and through the continuum limit of the topological susceptibility.

A central result is that the efficiency of O-REX can be predicted based on general theoretical arguments. The acceptance depends only on the average work done along the protocol, through exact relations that follow from Crooks' and Jarzynski's theorems, and the average work scales in a simple way with the number of replicas, the length of the protocol and the size of the defect. A standard REX simulation and a short scan in the protocol length are then enough to predict the acceptance of any setup and, together with a simple cost model, to fix the optimal protocol length, which agrees with the best setups found in our simulations.

The same scaling shows how to approach the continuum limit. As the defect grows, the acceptance can be kept fixed by adding replicas, by making the protocol longer, or both. This leaves a practical freedom: when memory limits the number of replicas, the protocol can be made longer, with a comparable gain. This remains true even when the protocol is too expensive to reduce the cost, for instance if the updates cannot be restricted to a small patch: at fixed acceptance, O-REX still requires fewer replicas, which can be the deciding factor when memory, rather than computing time, is the limit.

When simulating with the predicted optimal parameters, O-REX reduces the cost at fixed statistical precision by about $\gtrsim40\%$ with respect to standard REX, and this gain stays constant over the five lattice spacings we studied. The power-law scaling with the lattice spacing is instead unchanged: O-REX reduces the \emph{prefactor} of the cost. Part of the gain comes from the smaller number of replicas (O-REX only requires half of them to get the same acceptance rate as the standard REX), and part from the decorrelation induced by the local updates along the protocol. The gain comes almost for free, since any parallel tempering code already contains the local updates: O-REX only requires restricting them to a patch and accumulating the work along the protocol. 

The extension to $4d$ $\mathrm{SU}(N)$ gauge theories is straightforward, since PTBC has the same structure there: the action is linear in the defect coupling and the defect is localized. This is also where O-REX could be most useful. In four dimensions the defect is a three-dimensional region, so its volume in lattice units grows as $1/a^{3}$ towards the continuum limit, and the number of replicas required by standard REX as $1/a^{3/2}$. With O-REX, part of this growth can be moved to the protocol length. Moreover, since the work is a local property of the defect, $k$ and $\alpha$ could be determined in cheaper simulations on smaller volumes. In QCD each replica is also expensive in memory, which makes this freedom more valuable. On the other hand, restricting the updates to a patch is not straightforward with dynamical fermions, since the fermion determinant is non-local. Domain-decomposition methods~\cite{Luscher:2003qa,Luscher:2005rx}, and in particular the local factorization of the fermion determinant of Ref.~\cite{Ce:2016idq}, which allows the gauge field in a block of the lattice to be updated with the fields outside kept fixed, are a natural starting point toward a practical application of O-PTBC to simulations with dynamical fermions.

More generally, this work is a first step toward the use of other methods to replace the instantaneous swap with a more elaborate transformation, such as replica exchange combined with learned transport maps between replicas~\cite{Invernizzi:2022aaa,Zhang:2025aaa}, which has also been explored for SU(3) gauge theory~\cite{Abbott:2024mix}. Whatever the transformation, the cost analysis of Sec.~\ref{subsec:parameters} identifies what is needed for a gain: the acceptance must grow fast enough with the effort spent on the transformation, and the cost of the transformation must remain small compared with that of the MCMC updates of the whole system.

Stochastic normalizing flows~\cite{Caselle:2022acb,Bulgarelli:2024yrz,Bulgarelli:2024brv,Bonanno:2025pdp,Bonanno:2026zqy} are a particularly promising case. They combine out-of-equilibrium MCMC updates with trained deterministic layers, and in SU$(3)$ gauge theory they have been built with layers acting only near the defect~\cite{Bulgarelli:2024yrz,Bonanno:2025pdp}. This class of flows requires only a cheap training, its cost comes practically for free compared with MCMC updates of the whole volume, and it has been shown to improve on purely stochastic protocols on boundary conditions by about a factor of three in terms of cost reduction~\cite{Bonanno:2025pdp}. Layers of this kind could be inserted directly in the O-REX protocol, provided it remains symmetric. Since they come at practically no computational cost, any reduction of the work they provide would translate directly into a replica number reduction.

Finally, we would like to observe that both PTBC and O-PTBC mitigate topological freezing by reducing the dynamical critical exponent of $\mathcal{C}\,\tauint(Q^2)\propto\xi^z$. In the $2d$ theory studied in this work, the reduction is from $z \simeq 9$ down to $z \simeq 2$, see Sec.~\ref{subsec:continuum}. The latter result is in line with general theoretical expectations: $z=2$ is the critical exponent expected from a diffusive Monte Carlo dynamics, such as the one characterizing topological fluctuations in the open replica. Recent developments that leverage deep generative models to construct hierarchical multiscale samplers target precisely the problem of further reducing this type of critical slowing down~\cite{Abbott:2024knk,Bauer:2024byr,Singha:2025lsd,Singha:2026slc,Hasenfratz:2026azk,Hasenfratz:2026mae,Singha:2026qyx}, and their combination with O-REX is a possible direction for future work.

\acknowledgments
We thank Claudio Bonati, Massimo D'Elia, Andrea Giorgieri, Gurtej Kanwar and Satria Wydianto for useful discussions. A.~N., D.~P. and L.~V. acknowledge support by the Simons Foundation grant 994300 (Simons Collaboration on Confinement and QCD Strings). E.~C. acknowledges support by the European Union (ERC, FoundLatt, 101212401). Views and opinions expressed are however those of the author only and do not necessarily reflect those of the European Union or the European Research Council. Neither the European Union nor the granting authority can be held responsible for them. A.~N., D.~P. and L.~V. acknowledge support from the SFT Scientific Initiative of INFN. The work of D.~V.~is supported by STFC under Consolidated Grants No.~ST/X000680/1 and No.~UKRI/ST/C004237/1. A.~B. was supported by the Deutsche Forschungsgemeinschaft (DFG, German Research Foundation) as part of the CRC 1639 NuMeriQS -- project no.\ 511713970 and under Germany’s Excellence Strategy -- Cluster of Excellence ``Color meets Flavor'' (CmF) EXC 3107 -- 533766364. R.~D.~acknowledges funding from the European Union (EU) Next Generation EU (NGEU) -- National Recovery and Resilience Plan (NRRP) -- MISSION 4 COMPONENT 1, INVESTMENT N.4.1 -- CUP N.~I51J24000160007 (PhD Cycle XL, Ministerial Decree no.~629/2024). Numerical calculations have been performed on the \texttt{Leonardo} machine at Cineca, based on the agreement between INFN and Cineca, under project INF26\_npqcd.

The authors used Claude (Anthropic) and Codex (OpenAI) as assistants for drafting and editing parts of the text, for checking the manuscript for consistency, and in support of the analytical calculations, the data analysis and the figures. The simulation code was written entirely by the authors. All content was reviewed and verified by the authors, who take full responsibility for it.

\FloatBarrier

\appendix

\begin{table}[!b]
\centering
\footnotesize
\begin{tabular}{ccccccccccc}
\hline
$L$ & & $n_{\rm rep}$ & $n_{\rm step}$ & $L_\R$ & $\mathcal C$ & chains & MC steps & $\langle A\rangle$ (\%) & $\tau_{\rm int}(Q^2)$ & $\mathcal C\,\tau_{\rm int}(Q^2)$ \\
\hline
114 & REX & 7 & 0 & -- & 7.00 & 100 & 60k & 14.4 & 26.3(2.6) & 184(18) \\
114 & REX & 8 & 0 & -- & 8.00 & 100 & 183k & 21.1 & 24.2(1.7) & 194(14) \\
114 & REX & 14 & 0 & -- & 14.00 & 100 & 60k & 50.1 & 12.1(1.1) & 169(15) \\
114 & O-REX & 7 & 8 & 10 & 11.71 & 100 & 145k & 49.9 & 8.8(1.2) & 103(14) \\
114 & O-REX & 8 & 6 & 10 & 12.28 & 100 & 141k & 52.4 & 9.5(1.3) & 117(16) \\
\hline
150 & REX & 7 & 0 & -- & 7.00 & 200 & 95k & 8.8 & 40.2(1.9) & 281(13) \\
150 & REX & 8 & 0 & -- & 8.00 & 200 & 88k & 14.4 & 33.8(1.6) & 270(13) \\
150 & REX & 9 & 0 & -- & 9.00 & 200 & 61k & 20.1 & 27.4(1.5) & 247(13) \\
150 & REX & 14 & 0 & -- & 14.00 & 200 & 98k & 43.1 & 20.5(1.1) & 287(15) \\
150 & REX & 16 & 0 & -- & 16.00 & 200 & 91k & 49.5 & 18.3(1.1) & 292(18) \\
150 & O-REX & 6 & 2 & 13 & 7.26 & 200 & 75k & 14.2 & 34.1(1.7) & 247(13) \\
150 & O-REX & 6 & 3 & 13 & 7.68 & 200 & 71k & 19.2 & 30.9(2.0) & 237(15) \\
150 & O-REX & 7 & 2 & 13 & 8.51 & 200 & 64k & 22.1 & 28.5(1.7) & 243(14) \\
150 & O-REX & 7 & 3 & 13 & 9.02 & 200 & 60k & 27.6 & 23.0(1.7) & 207(15) \\
150 & O-REX & 7 & 8 & 13 & 11.54 & 200 & 114k & 42.7 & 17.0(1.6) & 196(18) \\
150 & O-REX & 7 & 10 & 13 & 12.54 & 200 & 108k & 46.1 & 15.4(1.2) & 194(15) \\
150 & O-REX & 8 & 8 & 13 & 13.29 & 200 & 104k & 49.6 & 12.9(1.1) & 172(14) \\
150 & O-REX & 8 & 10 & 13 & 14.47 & 200 & 98k & 52.7 & 12.5(1.0) & 180(15) \\
150 & O-REX & 7 & 8 & 50 & 59.84 & 200 & 67k & 43.3 & 10.6(1.3) & 634(80) \\
150 & O-REX & 7 & 10 & 50 & 71.59 & 200 & 64k & 46.8 & 8.95(95) & 641(68) \\
150 & O-REX & 8 & 8 & 50 & 69.65 & 200 & 62k & 50.1 & 7.67(75) & 534(52) \\
150 & O-REX & 8 & 10 & 50 & 83.35 & 200 & 59k & 53.3 & 6.54(83) & 545(69) \\
\hline
\end{tabular}
\caption{Main simulations at $L=114$ and $L=150$ ($\nstep=0$ $\to$ REX): $\mathcal C$ is the cost in Eq.~\eqref{eq:orex_cost} in units of full-lattice updating steps of one replica; $\tau_{\rm int}(Q^2)$ is in units
of measurements (one every 10 Monte Carlo steps); MC steps is the
length of each chain. All ladders use the optimal schedule.}
\label{tab:runs}
\end{table}

\section*{Appendix}

\section{Summary of the simulations}
\label{app:runs}
Tab.~\ref{tab:runs} lists the simulations at $L=114$ and $L=150$ used in Secs.~\ref{subsec:orex_scaling}--\ref{subsec:orex_cost}, and Tab.~\ref{tab:scaling} those at the other three lattice spacings of the line of constant physics, used in Sec.~\ref{subsec:continuum}.

\begin{table}[!t]
\centering
\footnotesize
\begin{tabular}{cccccccccc}
\hline
$L$ & & $n_{\rm rep}$ & $n_{\rm step}$ & $L_\R$ & $\mathcal C$ & MC steps & $\langle A\rangle$ (\%) & $\tau_{\rm int}(Q^2)$ & $\mathcal C\,\tau_{\rm int}(Q^2)$ \\
\hline
88 & REX & 14 & 0 & -- & 14.00 & 40k & 58.2 & 6.65(63) & 93(9) \\
88 & REX & 6 & 0 & -- & 6.00 & 40k & 15.3 & 15.6(3.0) & 94(18) \\
88 & REX & 7 & 0 & -- & 7.00 & 40k & 23.4 & 12.1(1.4) & 85(10) \\
88 & O-REX & 6 & 8 & 8 & 10.15 & 40k & 51.1 & 5.65(75) & 57(8) \\
88 & O-REX & 6 & 6 & 8 & 9.23 & 40k & 47.0 & 6.20(64) & 57(6) \\
88 & O-REX & 7 & 6 & 8 & 10.87 & 40k & 54.7 & 5.78(60) & 63(7) \\
\hline
102 & REX & 14 & 0 & -- & 14.00 & 55k & 54.0 & 10.65(72) & 149(10) \\
102 & REX & 6 & 0 & -- & 6.00 & 55k & 11.1 & 25.0(2.0) & 150(12) \\
102 & REX & 7 & 0 & -- & 7.00 & 55k & 18.4 & 21.7(1.7) & 152(12) \\
102 & O-REX & 6 & 8 & 9 & 9.94 & 55k & 46.2 & 8.5(1.2) & 84(11) \\
102 & O-REX & 7 & 8 & 9 & 11.73 & 55k & 53.9 & 6.78(82) & 80(10) \\
102 & O-REX & 6 & 7 & 9 & 9.51 & 55k & 44.2 & 9.2(1.1) & 87(10) \\
102 & O-REX & 7 & 7 & 9 & 11.21 & 55k & 52.1 & 9.6(1.4) & 108(16) \\
\hline
128 & REX & 14 & 0 & -- & 14.00 & 85k & 46.5 & 17.1(1.2) & 239(16) \\
128 & REX & 7 & 0 & -- & 7.00 & 85k & 11.4 & 32.0(1.6) & 224(11) \\
128 & REX & 8 & 0 & -- & 8.00 & 85k & 17.5 & 27.4(1.6) & 219(13) \\
128 & O-REX & 7 & 8 & 11 & 11.55 & 85k & 46.2 & 12.1(2.0) & 140(23) \\
128 & O-REX & 8 & 8 & 11 & 13.31 & 85k & 52.8 & 10.4(1.1) & 139(15) \\
128 & O-REX & 7 & 9 & 11 & 12.05 & 85k & 47.9 & 11.3(1.2) & 136(14) \\
128 & O-REX & 8 & 9 & 11 & 13.90 & 85k & 54.4 & 9.65(96) & 134(13) \\
\hline
\end{tabular}
\caption{Simulations at $L=88$, 102 and 128: REX simulations use either $n_{\rm rep}=14$ or the O-REX replica numbers, O-REX uses either $\nstep=8$ or $\nstep\simeq8\,L_\dd/7$. Same units as in Tab.~\ref{tab:runs}.}
\label{tab:scaling}
\end{table}

\FloatBarrier

\section{Choice of the patch size}
\label{app:local_updates}
In this appendix, we study the performance of O-REX upon varying the patch size $v=V_{\rm patch}/V$, cf.~Eq.~\eqref{eq:patch_volume}.

\begin{figure}[!t]
    \centering
    \includegraphics[width=\textwidth]{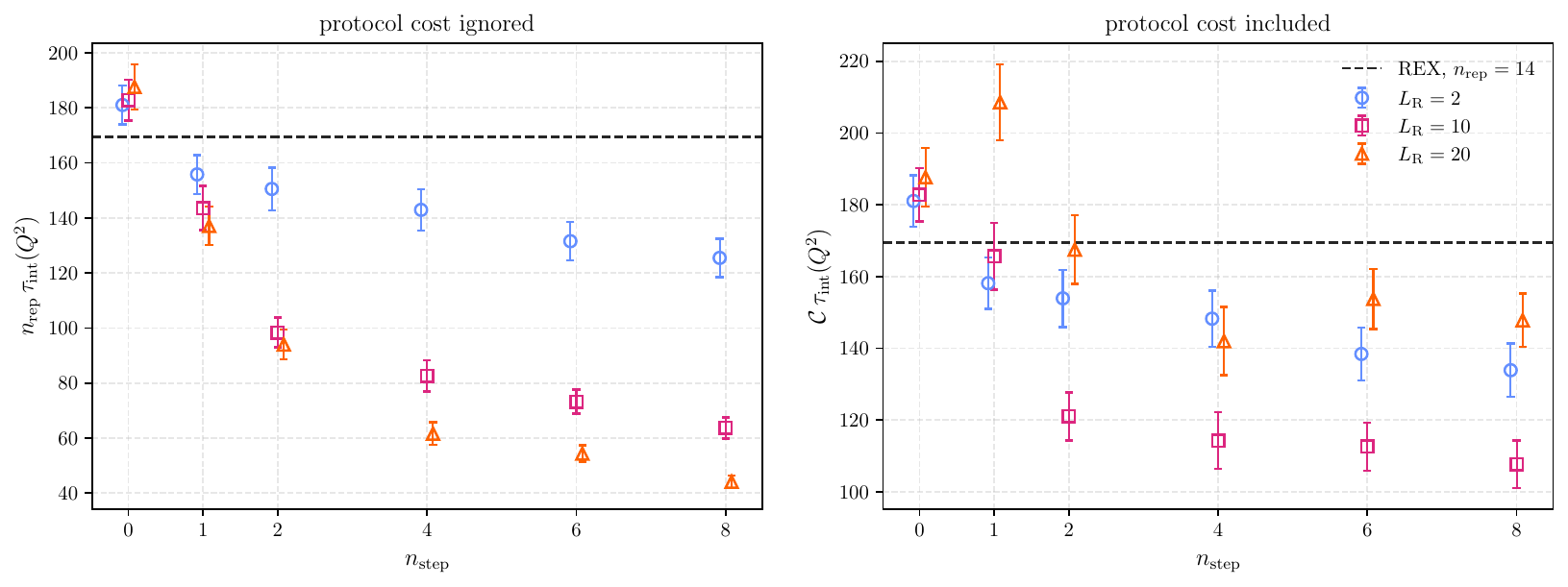}
    \caption{$L=114$: integrated autocorrelation time of $Q^2$ multiplied by the number of replicas (left, protocol cost omitted) and by the cost $\mathcal{C}$ of Eq.~\eqref{eq:orex_cost} (right), as a function of $\nstep$ for three patch sizes. Each point is the average over $\nrep=7$--$10$, using the optimal ladder. The dashed line is standard REX with $\nrep=14$ and approximately 50\% acceptance.}
    \label{fig:cost}
\end{figure}

Fig.~\ref{fig:cost} compares three patch sizes at $L=114$ as a function of the protocol length. When the protocol cost is omitted, the largest patch gives the lowest $\nrep\,\tauint(Q^2)$. Including the protocol cost, $L_\R=10\simeq1.8\,\xi/a$ gives the lowest $\mathcal{C}\,\tauint(Q^2)$ among the explored values of $v$ for every $\nstep\ge2$. Larger patches further reduce the autocorrelation time, but their additional cost outweighs this improvement.

\begin{figure}[!t]
    \centering
    \includegraphics[width=\textwidth]{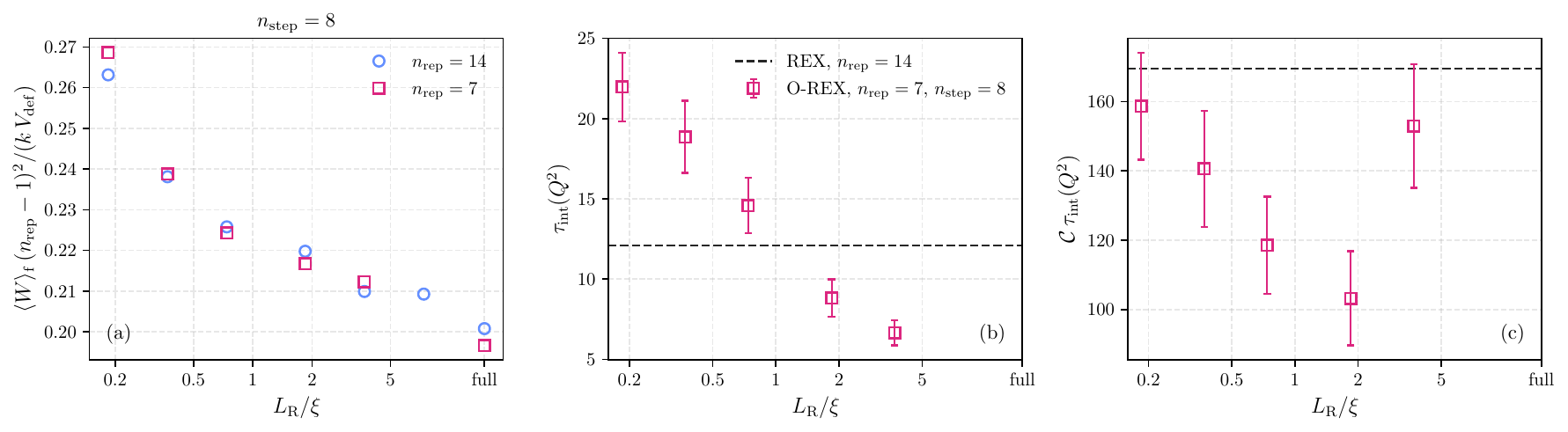}
    \caption{$L=114$, $\nstep=8$, optimal ladder, as a function of the patch half-size in units of $\xi$
(``full'': the whole lattice). Left: average work in units of $k\,V_{\mathrm{def}}/(\nrep-1)^2$, i.e.\ of the REX work given by Eq.~\eqref{eq:work_scaling} with the value of $k$ of Fig.~\ref{fig:scaling}, for
$\nrep=14$ and 7. Center and right: $\tauint(Q^2)$ and
$\mathcal{C}\,\tauint(Q^2)$ for $\nrep=7$; the dashed line is standard REX
with $\nrep=14$.}
    \label{fig:patch114}
\end{figure}

\begin{figure}[!t]
    \centering
    \includegraphics[width=\textwidth]{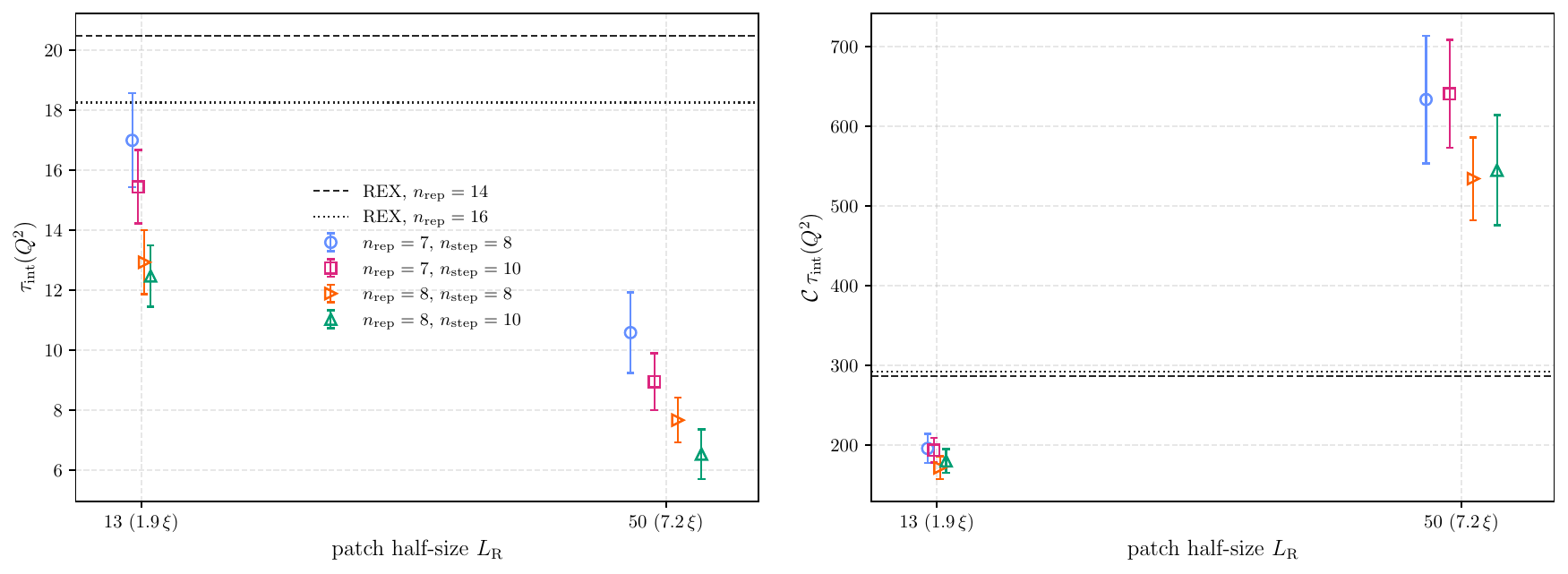}
    \caption{$L=150$: $\tauint(Q^2)$ (left) and $\mathcal{C}\,\tauint(Q^2)$ (right) for four O-REX setups with patch half-size $L_\R=13$ ($1.9\,\xi$) and $L_\R=50$ ($7.2\,\xi$, 49\% of the lattice volume). The points at each patch size are slightly displaced horizontally for readability. Dashed and dotted lines show standard REX with $\nrep=14$ and $16$, respectively.}
    \label{fig:patch150}
\end{figure}

Fig.~\ref{fig:patch114} shows the $L_\R$-dependence of several quantities at $L=114$ and $\nstep=8$. The work
keeps decreasing as the patch grows, from $L_\R=1$ to $L_\R=L/2$. The patch has a much larger effect on $\tauint(Q^2)$ than on the
acceptance: between $L_\R=1$ and $L_\R=20$, $\tauint$ falls by a factor 3.3, while the
acceptance only rises from 45\% to 50\%. Since the cost grows as $L_\R^2$, $\mathcal{C}\,\tauint$ has a minimum at $L_\R\simeq2\xi/a$, as also found numerically at $L=150$ (see Fig.~\ref{fig:patch150}).

Fig.~\ref{fig:patch150} compares the patches $L_\R=13$ and $L_\R=50$ at $L=150$, for $\nrep=7,8$ and $\nstep=8,10$. The larger patch covers about 49\% of the lattice volume and reduces $\tauint(Q^2)$ by a factor of $1.6$--$1.9$, while the acceptance changes by less than one percentage point. The cost, however, increases by a factor of $5$--$6$, so that $\mathcal{C}\,\tauint(Q^2)$ becomes roughly twice that of standard REX. For all four setups, the $L_\R\sim 2\xi/a$ patch gives again the best efficiency.

The position of this minimum depends on how the cost of the patch updates is counted. In our implementation the updates are sequential, and the cost of a sweep is proportional to the number of updated sites, which is the assumption behind Eq.~\eqref{eq:orex_cost}. With a parallel implementation, e.g.\ a checkerboard (even--odd) update on GPUs, the cost of updating a small patch does not decrease proportionally to its volume, so that the relative cost of the protocol is larger than $V_{\mathrm{patch}}/V$. The optimal patch size, and with it the optimal length of the protocol, then depend on the implementation, and should be determined with the corresponding cost model.

\section{Optimal ladder}
\label{app:optimal_protocol}
The defect couplings of the replica ladder are chosen following Refs.~\cite{Syed:2021nrpt,Syed:2024}. The local metric of the ladder is the variance of the defect action, $\mathrm{Var}_c(S_{\L}^{{\sst{(\mathcal{D})}}})$,
measured at each coupling of a standard REX simulation with a linear ladder of 14 replicas
(100 chains, first fifth of each chain discarded). Integrating its square root with the trapezoidal
rule gives the cumulative function $\Lambda(c)=\int_0^c \ddiff c^\prime\,\sqrt{\mathrm{Var}_{c^\prime}(S_{\L}^{\sst{(\mathcal{D})}})}$ and the thermodynamic length:
\begin{equation}
    \Lambda\equiv \Lambda(1)=\int_0^1 \ddiff c\,\sqrt{\mathrm{Var}_c(S_{\L}^{\sst{(\mathcal{D})}})}.
    \label{eq:thermo_length}
\end{equation}

\begin{figure}[!t]
    \centering
    \includegraphics[width=\textwidth]{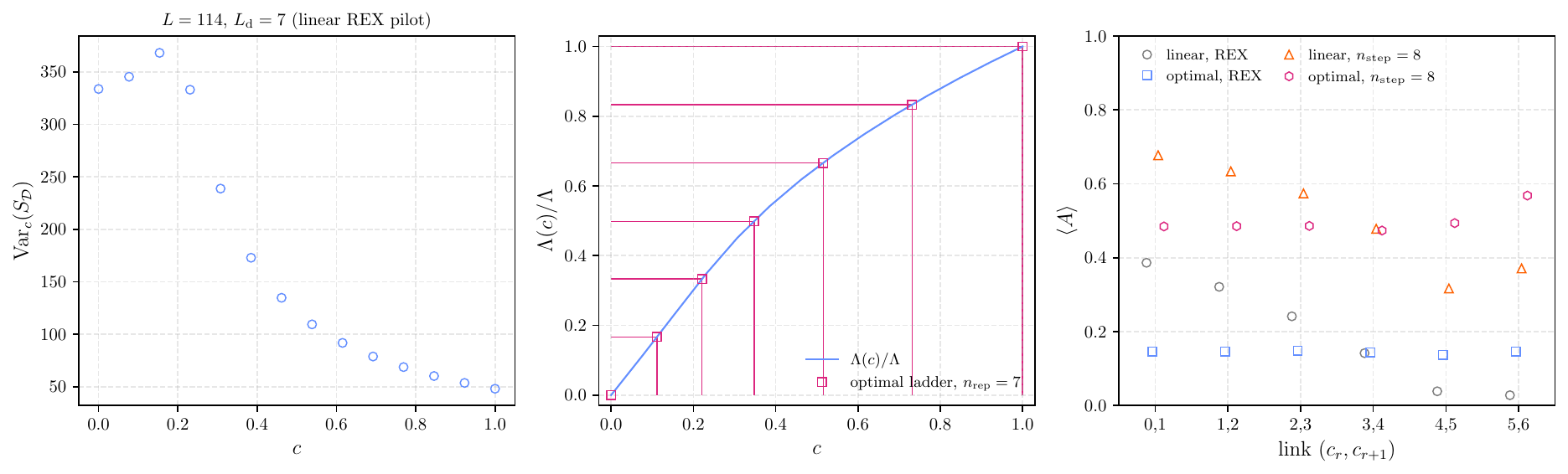}
    \caption{Optimal ladder at $L=114$, $L_\dd=7$. Left: variance of the defect action measured with a linear
ladder of 14 replicas. Center: normalized cumulative function $\Lambda(c)/\Lambda$ and the
optimal couplings for $\nrep=7$, evenly spaced in $\Lambda(c)$. Right: acceptance on
each pair $(c_i,c_{i+1})$ for $\nrep=7$, with linear and optimal ladders, for REX and for O-REX with
$\nstep=8$, $L_\R=10$.}
    \label{fig:schedule}
\end{figure}

The optimal ladder with $\nrep$
replicas is obtained by inverting $\Lambda(c)$, through a cubic spline, at
$\Lambda(c_i)=\Lambda\,(\nrep-1-i)/(\nrep-1)$. On this ladder every link
has the same REX acceptance, $1-\mathrm{erf}[\Lambda/(2(\nrep-1))]$. At $L=114$ the
measured ladder average reproduces this analytic result within 0.05 percentage points for $\nrep=7$ and
14 (Fig.~\ref{fig:schedule}). The same ladder is
used for O-REX, where it again equalizes the acceptance of the bulk replicas. Only the replica next to the
open end has a slightly higher acceptance, since there the defect relaxes faster along the
protocol. The values of $\Lambda$ along the line of constant physics are listed in Tab.~\ref{tab:lambda}.

\begin{table}[!t]
\centering
\begin{tabular}{cccc}
\hline
$L$ & $\beta$ & $\Lambda$ & $\langle A_{\rex}\rangle_{14}$ (\%) \\
\hline
88 & 0.660 & 10.120(3) & 58.2 \\
102 & 0.680 & 11.280(4) & 54.0 \\
114 & 0.700 & 12.391(3) & 50.0 \\
128 & 0.718 & 13.429(5) & 46.5 \\
150 & 0.741 & 14.481(5) & 43.1 \\
\hline
\end{tabular}
\caption{Thermodynamic length $\Lambda=\int_0^1 \ddiff c\,\sqrt{\mathrm{Var}_c(S_{\L}^{\sst{(\mathcal{D})}})}$ of the defect coupling along the line of constant physics of Tab.~\ref{tab:physical}, measured on 100 independent standard REX chains with a linear ladder of 14 replicas. Here $\langle A_{\rex}\rangle_{14}=1-\mathrm{erf}[\Lambda/(2\cdot 13)]$ is the exchange acceptance of standard REX with the optimal ladder of $n_{\rm rep}=14$ replicas.}
\label{tab:lambda}
\end{table}

\section{Work distributions}
\label{app:work_distributions}
\begin{figure}[!htbp]
    \centering
    \includegraphics[width=\textwidth]{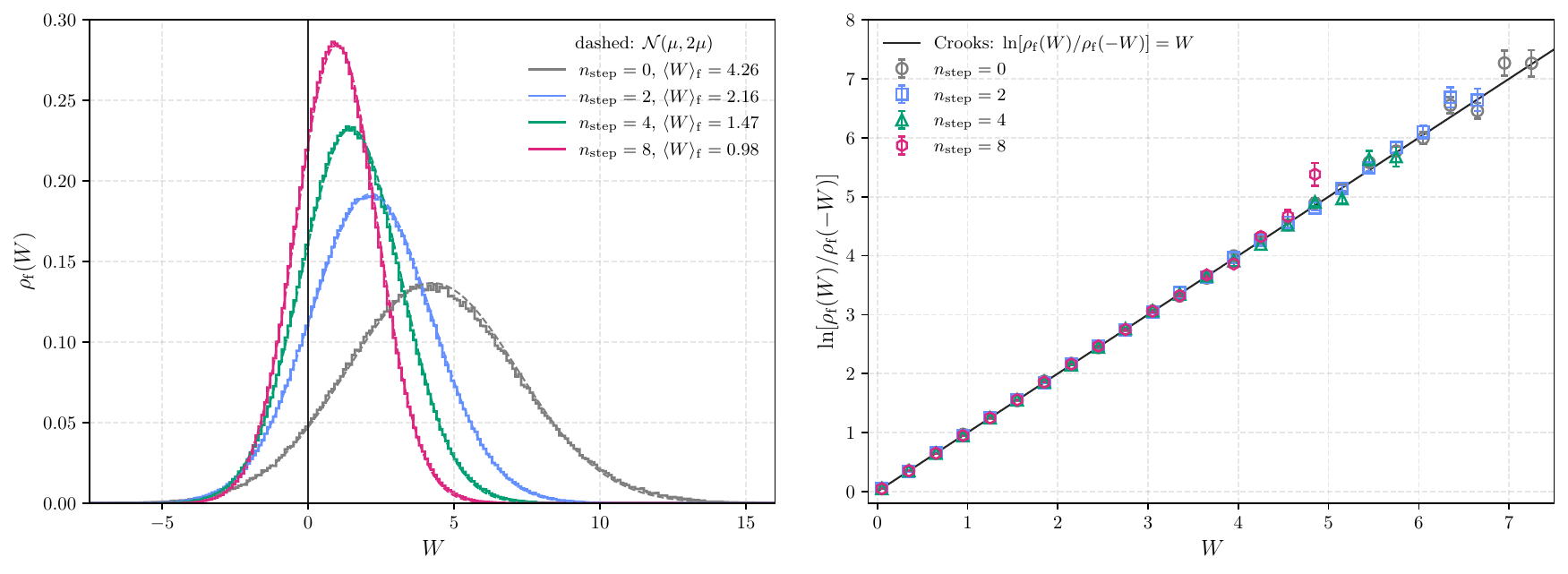}
    \caption{Left: distribution of the work on the link $(c_2,c_3)$ of the optimal ladder with
$\nrep=7$ at $L=114$, $L_\R=10$, for $\nstep=0$ (REX), 2, 4, 8; the dashed curves are
Gaussians with the measured mean $\mu$ and variance $2\mu$. Right: Crooks relation,
$\ln[\rho_f(W)/\rho_f(-W)]=W$, for the same protocols, using all six links of the ladder; the
reverse protocol is the forward one with the replicas relabeled, so $\rho_r=\rho_f$.
$(0.6$--$1.7)\times10^6$ trajectories per link.}
    \label{fig:workdist}
\end{figure}

Fig.~\ref{fig:workdist} shows the full distribution of the work for one pair $(c_i,c_{i+1})$ of the optimal ladder. For every $\nstep$, it is Gaussian with $\sigma_W^2=2\langle W\rangle_    f$, Eq.~\eqref{eq:gaussian_jarzynski}, and it satisfies the Crooks relation $\rho_\f(W)=\ee^W\rho_\f(-W)$, Eq.~\eqref{eq:crooks_work} with $\rho_\r=\rho_\f$, up to $W\simeq7$, i.e., for a range of values of $\rho_\f$ spanning three orders of magnitude.

\section{Performance comparison at fixed swap rate}
\label{app:acceptances}

\begin{figure}[!t]
    \centering
    \includegraphics[width=0.75\textwidth]{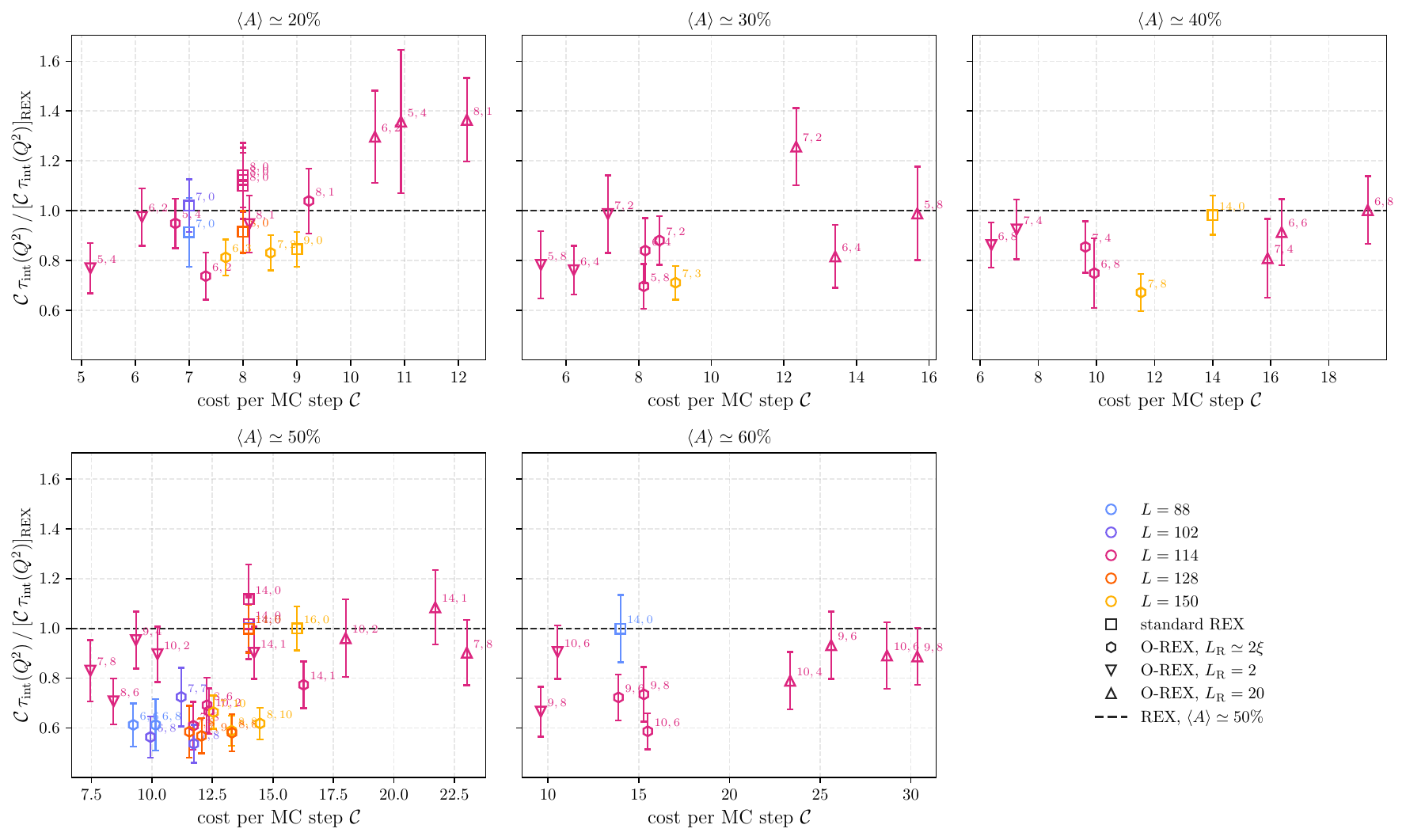}
    \caption{Cost measure, relative to standard REX with acceptance close to $50\%$ on the same lattice, against the cost per Monte Carlo step, for all setups whose mean exchange acceptance lies within 2.5 ($L=114$) or 4 (other lattices) percentage points of 20, 30, 40, 50 and 60\%. Colors denote the lattice and markers the algorithm and the patch size; labels give $\nrep,\nstep$, with $\nstep=0$ for standard REX. Lower is better.}
    \label{fig:isoacc}
\end{figure}

\begin{figure}[!t]
    \centering
    \includegraphics[width=\textwidth]{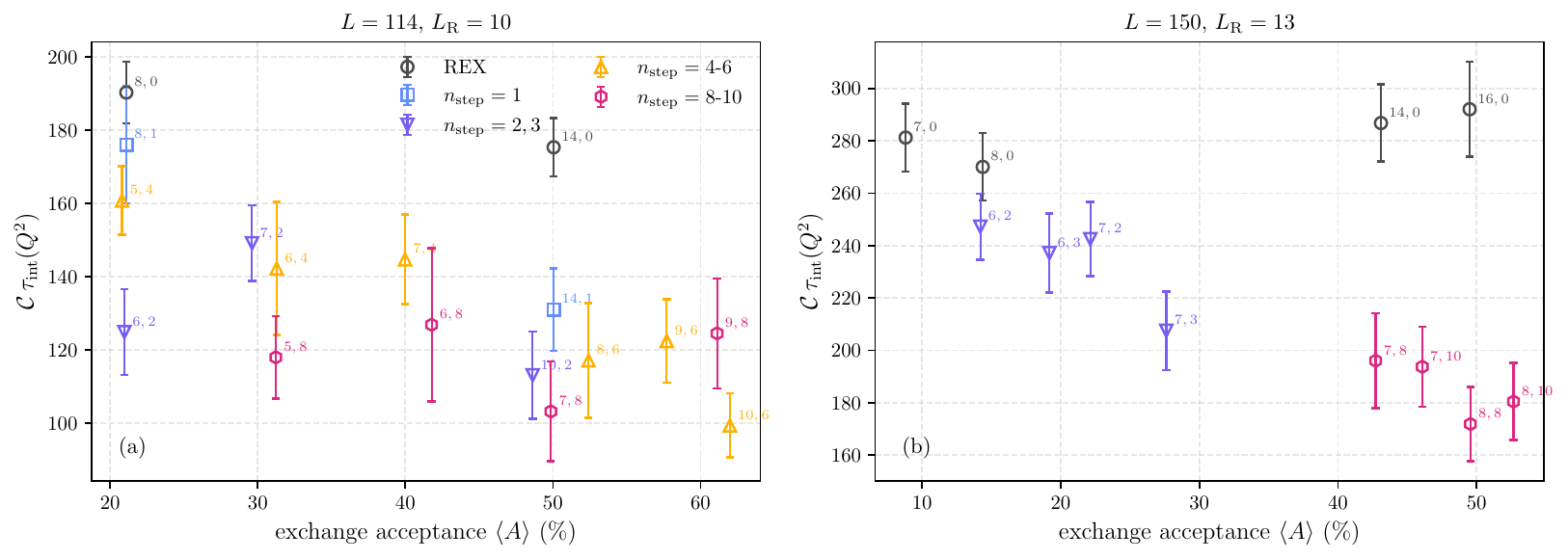}
    \caption{Cost measure against the exchange acceptance at $L=114$ (a) and $L=150$ (b), for standard REX and for O-REX with $L_\R=10$ and $L_\R=13$, respectively, grouped by protocol length; labels give $\nrep,\nstep$. At $L=114$, the standard REX runs repeated in the three patch scans are combined into their weighted mean.}
    \label{fig:fomacc}
\end{figure}

Since increasing the swap rate in O-REX can be done by increasing either $\nrep$, $\nstep$, or $v$, the cost of different algorithmic setups is compared at fixed acceptance. Fig.~\ref{fig:isoacc} shows the figure of merit $\mathcal{C} \tauint(Q^2)$ for all setups, at the five lattice spacings, whose acceptance lies within 2.5 ($L=114$) or 4 (other lattices) percentage points around 20, 30, 40, 50 and 60\%. The O-REX cost measure reported there is given in terms of the standard REX one with acceptance close to $50\%$ on the same lattice. In the end, for every explored acceptance rate and lattice spacing, O-REX setups with $L_\R\simeq2\xi/a$ give the lowest $\mathcal{C}\,\tauint(Q^2)$. In Fig.~\ref{fig:fomacc} we show the cost measure as a function of the acceptance. For standard REX it is essentially flat in the explored range of swap rates, while for O-REX it decreases as a function of the acceptance rate, and is lowest between 50\% and 60\%. As an example, at those values of $\braket{A_{\orex}}$, the optimal setup at $L=114$ gives $\mathcal{C}\,\tauint=103(14)$ for 50\% and $99(9)$ for 62\%.

\section{Non-topological observable comparison with standard REX}
\label{app:exactness}

\begin{figure}[!t]
    \centering
    \includegraphics[width=0.75\textwidth]{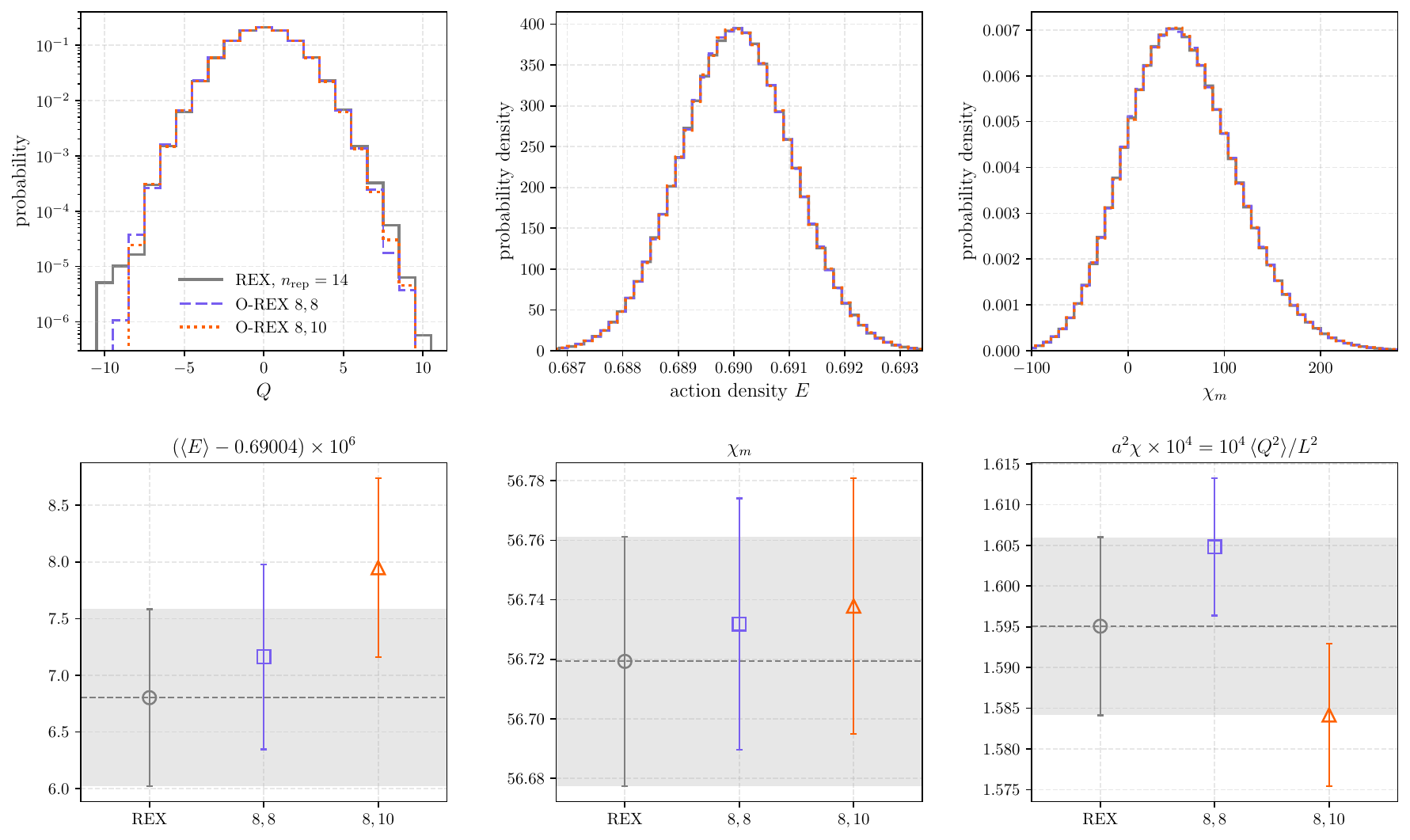}
    \caption{The plot compares standard REX ($\nrep=14$) against O-REX with $\nrep=8$, $\nstep=8,10$, $L_\R=13$. Results refer to $L=150$. Top panels: distributions of the topological charge, of the action density, and of the magnetic susceptibility estimator. Bottom panels: the corresponding averages (the band corresponds to the standard REX data).}
    \label{fig:exactness}
\end{figure}

Since O-REX is exact for any $\nstep$, every equilibrium observable on the periodic replica must agree with standard REX at the same lattice spacing. We want to check this at high precision for the path-integral distribution of several observables: the action density $E$, the magnetic susceptibility $\chi_{_{\sst{\rm m}}}$, and the topological charge $Q$. In Fig.~\ref{fig:exactness} we compare results obtained with the two best $L=150$ O-REX setups, and with standard REX (in all cases using $ \sim 2 \times10^6$ measurements). The expectation values of the action density, the magnetic susceptibility, and the topological susceptibility agree within at most one standard deviation. Also the probability distributions of $Q$, $E$ and of the magnetic susceptibility perfectly agree among the different algorithms: the
$\tilde{\chi}^2/\mathrm{dof}$ between the histograms is
0.83--0.85 for $Q$, 0.71--1.02 for $E$ and 0.62--0.69 for the magnetic susceptibility estimator.

\FloatBarrier
\bibliographystyle{JHEP}
\bibliography{biblio.bib}

\end{document}